\documentclass[preprint]{aastex631}

\begin{document}

\title{A broad set of solar and cosmochemical data indicates high C-N-O abundances for the solar system}

\author[0000-0003-2689-3102]{Ngoc Truong*}
\author[0000-0002-2161-4672]{Christopher R. Glein}
\affiliation{Space Science Division, Southwest Research Institute \\
6220 Culebra Rd, San Antonio, TX 78238, USA}

\author[0000-0003-2279-4131]{Jonathan I. Lunine}
\affiliation{Department of Astronomy \& Carl Sagan Institute, Cornell University \\
122 Sciences Dr, Ithaca, NY 14850, USA }

\correspondingauthor{Ngoc Truong}
\email{ngoc.truong@swri.edu}



\begin{abstract}

We examine the role of refractory organics as a major C carrier in the outer protosolar nebula and its implications for the compositions of large Kuiper belt objects (KBOs) and CI chondrites. By utilizing Rosetta measurements of refractory organics in comet 67P/Churyumov-Gerasimenko, we show that they would make up a large fraction of the protosolar C inventory in the KBO-forming region based on the current widely adopted solar abundances. However, this would free up too much O to form water ice, producing solid material that is not sufficiently rock-rich to explain the uncompressed density of the Pluto-Charon system and other large KBOs; the former has been argued as the most representative value we have for the bulk composition of large KBOs (Barr \& Schawmb 2016, Bierson \& Nimmo 2019). This inconsistency further highlights the solar abundances problem - an ongoing challenge in reconciling spectroscopically determined heavy element abundances with helioseismology constraints. By employing a new dataset from solar CNO neutrinos and solar wind measurements of C, N, and O, we show that the uncompressed density of the Pluto-Charon system can be reproduced over a wide range of scenarios. We show that a lack of sulfates in Ryugu and Bennu samples implies a lower amount of water ice initially accreted into CI chondrite parent bodies than previously thought. These data are found to be consistent with the solar C/O ratio implied by the new dataset. Our predictions can be tested by future neutrino, helioseismology, and cosmochemical measurements. 

\end{abstract}



\section{Introduction} \label{sec:intro}
The Sun’s heavy element abundances play a crucial role in numerous areas of astronomy. For example, they serve as benchmarks for understanding other stars’ elemental compositions (Hinkel et al. 2014; Jofré et al. 2019). Though the abundances of heavy elements (all elements other than H, He)  in the Sun are believed to be known to much better accuracy than in other stars, significant uncertainties remain (Basu \& Antia 2008). Traditionally, spectroscopic observations of the solar photosphere have been used to estimate the abundances of heavy elements in the top layer of the Sun following the pioneering works of Russell 1929 and Goldberg et al. 1960. However, the analysis of spectral lines to derive solar abundances is subject to uncertainties stemming from radiative transfer models, overlapping of spectral lines between different elements and elemental opacities within the Sun’s interior, among other complications (Jofré et al. 2019). We refer the reader to Hinkel et al. 2014; Allende Prieto 2016; Jofré et al. 2019 for further details on the nuances in determining stellar abundances.  \\

Early spectral analyses often employed 1D, local thermodynamic equilibrium (LTE) radiative transfer models (Anders \& Grevesse 1989 - AG89; Grevesse \& Sauval 1998 - GS98), which despite their simplicity, agreed well with constraints inferred from helioseismology, such as the base of the convective zone (\(R_{CZ}\)), the sound speed and density profiles, and the He abundance at the top surface of the Sun (\(Y_S\)) (Basu \& Antia 2008). Since 2005, the introduction of 3D models has significantly decreased the estimated abundances for several key elements (i.e., C, O), thus the relative mass fraction of heavy elements with respect to hydrogen \textit{Z/X} to $0.0165 \pm 0.0011$ (Asplund et al. 2005 - AGS05) (here, \textit{X, Y, Z} are the mass fractions of H, He, and heavy elements, respectively). Monte Carlo simulations involving 10,000 solar models that utilize the solar abundances determination from AGS05 predict a surface He abundance \(Y_S\) = 0.2292 $\pm$ 0.0037 and \(R_{CZ}\) = (0.7280 $\pm$ 0.0037)$\times$\(R_\odot\) (\(R_\odot\): the radius of the Sun) (Bahcall et al. 2005, 2006). These results, however, are inconsistent with observations from helioseismology, in which $(Y_S\) = $0.2485 \pm 0.0034$, \(R_{CZ}\) = ($0.7133 \pm 0.0005$)$\times$\(R_\odot\), Basu \& Antia 2004, 2008). In contrast, those observations can be reproduced in solar models with the older, higher heavy element abundances data from AG89 or GS98, i.e., $Y_S\ = 0.2425 \pm 0.0042$ and \(R_{CZ}\) = ($0.7133 \pm 0.0035$)$\times$\(R_\odot\)  in models using GS98 data (Basu \& Antia 2008). Subsequent 3D solar models by Asplund et al. 2009 – AGSS09 and Asplund et al. 2021 – AGS21 slightly revised the \(Z/X\) value upward to be in the range of $0.0187 \pm 0.0009$, but this range is still well below older determinations from AG89 ($0.0274 \pm 0.0016$) or GS98 ($0.0231 \pm 0.0018$). While the disagreement in the more recent datasets is less severe than in AGS05, they are still in conflict with helioseismic data (Serenelli et al. 2009; Vinyoles et al. 2017). \\

Numerous attempts have been made to reconcile the low-$Z$ solar models with helioseismic data (Christensen-Dalsgaard et al. 2009; Serenelli et al. 2011; Bailey et al. 2015); however, they often improve some constraints but sacrifice their consistency with others (Haxton et al. 2013; Vinyoles et al. 2017). Among those ideas, opacity adjustments of certain heavy elements (e.g., iron) have been proposed based on transmission opacity experiments as one of the most promising potential solutions to resolve the problem (Bailey et al. 2015; Nagayama et al. 2019). However, a recent radiation burn-through experiment does not support a large increase in the opacity of iron at conditions close to the base of the solar convective zone and suggests that it agrees with current predictions from plasma opacity theory  (Hoarty et al. 2023). Alternatively, it has been suggested that the accretion of metal-poor disk gases (due to planetary formation) onto the Sun’s convective envelope might dilute its protosolar composition (Haxton et al. 2013; Kunitomo \& Guillot 2021; Kunitomo et al. 2022). While it is an interesting idea attempting to reconcile spectroscopic, helioseismic and neutrino constraints, challenges remain (Haxton et al. 2013). As an example, the timescale for the presence of gas in the protoplanetary disk is typically estimated to be a few Myrs (Wang et al. 2017), whereas the early Sun’s convective boundary moves outward in response to the growing radiative zone over a longer period of $ \sim 30$ Myr (Haxton et al. 2013). As a result, if the accretion of metal-poor disk gases occured early when most of the solar envelope was still convective, any effect would have been negligible (Haxton et al. 2013). Recent developments in 3D, non-LTE models, such as those by Magg et al. 2022, have revised upwards the abundances of elements like C, N, and O. Nevertheless, further efforts are still needed to reconcile these results with helioseismology observations (Buldgen et al. 2023). \\

Another source of valuable information on solar abundances is CI carbonaceous chondrites (Cameron 1973, 1982; Lodders 2003). The abundance pattern of many elements, except the most volatile ones (i.e., C, N, O, noble gases), show a close correlation with photospheric abundances (Lodders 2021). However, as Lodders 2020, 2021 has pointed out and quantified, heavy elements have gravitationally settled from the outer convective zone into the Sun’s interior over the lifetime of the Sun; therefore, the bulk composition of the Sun (and thus the rest of the solar system abundances), or protosolar abundances on the so-called astronomical scale (relative to $10^{12}$ hydrogen atoms), needs to be corrected from the contemporary photospheric abundances. Since the effect is not resolvable for individual elements with current data, in practice, the typical approach that has been taken is to adopt a constant settling factor (usually $\sim$ 10-23 \%) for all heavy elements (except Li). Note, however, that on the cosmochemical scale (normalized to $10^6$ silicon atoms), the contemporary abundances of heavy elements stay the same as the protosolar abundances (Lodders 2020, 2021). \\

In planetary and exoplanetary sciences, the abundances of heavy elements relative to the protosolar values are used as a diagnostic set to probe the formation conditions of planets (Owen et al. 1999; Öberg et al. 2011, Mousis et al. 2021). In particular, the solar C/O ratio and the resulting speciation of O among various forms of C (i.e., $CO$, refractory organics or organic matter, etc.) influences the composition of pebbles/planetesimals, thereby affecting the accretion of volatiles that can be represented in present planetary atmospheres (Pekmezci et al. 2019). The signatures of protosolar nebula chemistry are also likely to have left marks on more primitive bodies that formed in the outer solar system. These bodies include CI chondrites (Kruijer et al. 2017; Desch et al. 2018; Alexander 2019) and Kuiper belt objects (KBOs), leftovers of the era of planet formation. A more oxidized solar nebula, in which C was predominately present as $CO$ and $CO_2$, has often been invoked to account for the relatively low fraction of water ice initially accreted into CI chondrite parent bodies ($\sim$ 28-30 \% by weight, Alexander 2019). However, this constraint needs to be revisited, as analyses of returned samples from asteroid Ryugu and early data from asteroid Bennu both indicate the apparent absence of sulfates and suggest that CI sulfates are artifacts from the oxidation of indigenous metal sulfides in air (Nakamura et al. 2023; Yokoyama et al. 2023; King et al. 2024; Lauretta et al. 2024). This finding suggests that previous estimates of water ice content initially accreted into CI chondrites based on the presence of sulfates (e.g., Alexander 2019) may represent an upper limit. In light of this new insight, the implications to the protosolar nebula chemistry and protosolar abundances will be reassessed in section 3.2. \\

The compositions of KBOs could serve as another record of conditions in the early outer solar system. Their densities may reflect the amount of water ice that was present. There may have been a common grain density for mixtures of solids in the KBO-forming region. In this case, while KBOs come in a range of sizes and masses, porosity would explain the density variations between small and large KBOs (Bierson \& Nimmo 2019). Some small, low-density KBOs may be fragments of a differentiated parent body, as it was suggested for collisional families like Eris-Dysnomia (Brown et al. 2007). Alternatively, this trend could also be produced by the pebble accretion and streaming instability model if accreted pebbles were relatively rich in silicates (Cañas et al. 2024). The large KBOs are generally rock-rich with a canonical rock mass fraction of 70 \% (Nimmo et al. 2017; Ortiz et al. 2017; Bierson \& Nimmo 2019; Kiss et al. 2019); they have often been referred to as a representative composition of the solid materials in the outer protosolar nebula (McKinnon \& Mueller 1988; Simonelli et al. 1989; McKinnon et al. 2017, 2021). Furthermore, KBOs that were formed via low-velocity collisions, with impact speeds similar to escape speeds, would make large primary/moon companions similar to Pluto/Charon or Orcus/Vanth, and should retain their primordial compositions (Canup 2005; Barr \& Schwamb 2016; Canup et al. 2021; McKinnon et al. 2021). In contrast, collisional families like Eris-Dysnomia could be subjected to more energetic collisions that strip away their outer ice shells, leading to an increase in their rock mass fraction and bulk density (Brown et al. 2007; Barr \& Schwamb 2016; McKinnon et al. 2021). Neptune’s moon Triton is believed to be a captured Kuiper belt object (Goldreich et al. 1989; McKinnon \& Leith 1995; Agnor \& Hamilton 2006) with an uncompressed density $\sim$ 1900 kg $m^{-3}$, more rock-rich even than the Pluto-Charon system (McKinnon et al. 1995, Wong et al. 2008).  The capture process may have altered its original volatile composition (Goldreich et al. 1989); and subsequent tidal heating could have modified a more ice-rich primordial Triton, potentially from an original Pluto-like density ($\sim$ 1800 kg $m^{-3}$) to the current density (Barr \& Schwamb 2016). \\

Thus, the uncompressed density of the Pluto-Charon system, which was precisely measured by the New Horizons mission ($1800 \pm 20$ kg $m^{-3}$, 3$\sigma$ error bar, Brozović et al. 2015, McKinnon et al. 2017, Nimmo et al. 2017) has been argued as the most representative value we have for the bulk composition of large KBOs (Barr \& Schwamb 2016, Bierson \& Nimmo 2019, McKinnon et al. 2021). The rock-rich nature of KBOs has been commonly cited as evidence signifying their formation in a CO-rich protosolar nebula (McKinnon \& Mueller 1988; Johnson \& Lunine 2005; Wong et al. 2008), assuming a solar C/O ratio of $\sim$ 0.5, close to the current widely adopted value from AGSS09 ($\sim$ 0.54). While some previous works considered refractory organics as one of the C carriers in the protosolar nebula (Simonelli et al. 1989; Pollack et al. 1994) that could be inherited by icy moons (Miller et al. 2019; Néri et al. 2020; Reynard \& Sotin 2023), quantitative calculations of their effects have not yet been performed in detail for KBOs. As Jupiter-family comets are thought to originate in the Kuiper belt (Levison \& Duncan 1997; Volk \& Malhotra 2008; Nesvorný et al. 2017), their solid organic abundances may reflect the inventory of refractory organics in the KBO-forming region of the protosolar nebula. The Rosetta mission to comet 67P/Churyumov-Gerasimenko (67P/C-G) – a Jupiter-family comet provided a unique opportunity to sample such primordial materials at low flyby/orbital speeds ($<$ 10 m $s^{-1}$), likely preserving their original refractory compositions (Fray et al. 2016; Bardyn et al. 2017). In light of the new data from comet 67P/C-G, in the next section, we perform mass balance calculations to quantify the effects of solid organics on the uncompressed densities of KBOs under the current widely adopted solar abundances reported in Lodders 2021 (Table 1). \\

\begin{tabular}{ |c|c|c|c|c|}
 \hline

\ {Abundances} & \multicolumn{2}{|c|}{Present} & \multicolumn{2}{|c|}{Protosolar}\\
\hline

\ {Element} & N(E) & $\pm\sigma$&N(E) & $\pm\sigma$\\
\hline
 H   & $3.09\times 10^{10}$   &&   $2.52\times 10^{10}$ & \\

 He & $2.59\times 10^{9}$ & $1.2\times 10^{8}$ & $2.51\times 10^{9}$ & $1.2\times 10^{8}$ \\

 C& $9.12\times 10^{6}$ & $1.8\times 10^{5}$ &$9.12\times 10^{6}$& $1.8\times 10^{5}$\\
N &$2.19\times 10^{6}$ & 9700&  $2.19\times 10^{6}$ & 9700\\
O & $1.66\times 10^{7}$ & $1.3\times 10^{5}$ & $1.66\times 10^{7}$ & $1.3\times 10^{5}$ \\
Mg & $1.03\times 10^{6}$ & $4.4\times 10^{4}$ & $1.03\times 10^{6}$ & $4.4\times 10^{4}$ \\
Si & $1.00\times 10^{6}$ & $3.4\times 10^{4}$ & $1.00\times 10^{6}$ & $3.4\times 10^{4}$ \\
Al & $81820$ & $6110$ & $81820$ & $6110$ \\
Ca & $57239$ & $4500$ & $57234$ & $4500$ \\
Ni & $48670$ & $2940$ & $48670$ & $2940$ \\
Fe & $8.72\times 10^{5}$ & $3.8\times 10^{4}$ & $8.72\times 10^{5}$ & $3.8\times 10^{4}$ \\
S  & $4.37\times 10^{5}$ & $2.6\times 10^{4}$ & $4.37\times 10^{5}$ & $2.6\times 10^{4}$ \\
 \hline
\end{tabular}

\textbf{Table 1}. The current values adopted by the community for the present (left) and protosolar (solar system, right) abundances of major elements that determine the nature of solids formed in the outer solar system (Lodders 2021). These values are based on the cosmochemical abundance scale where the number of silicon atoms is set to one million ($n_{Si}$ = $10^6$). The present paper reexamines values for carbon, nitrogen, and oxygen (see Table 2). \\

\section{KBOs uncompressed densities and their formation: Insights from cometary data} \label{sec:style}

To assess the effects of C and O protosolar abundances on the compositions of KBOs, their uncompressed densities are calculated as a function of carbon partitioning (Equation 1, below). Uncompressed densities reflect the densities of the mixture of materials that make up a planetary body, after adjusting for the effect of gravity on the observed density. We model the density here, as it is the most reliable observable that is available for KBOs. We assume that the major C carriers in the outer protosolar nebula were $CO$, $CO_2$ and refractory organics ($C_{100}H_{104}O_{30}N_{3.5}$), the latter were measured in comet 67P/C-G’s dust (Bardyn et al. 2017; Isnard et al. 2019).\\

On the cosmochemical scale (relative to silicon, Table 1), the molar abundance of refractory organics in the KBO-forming region ($n_{C-org}$) is constrained by the ratio of C/Si measured in comet 67P/C-G’s dust ($5.5_{-1.2}^{+1.4}$ , Bardyn et al. 2017). The C/Si ratio in comet Halley (C/Si $\sim$ 5) is similar, although the flybys of the Vega-1, Vega-2 and Giotto spacecraft were performed at much higher speeds (Kissel et al. 1986; Kissel \& Krueger 1987; Jessberger et al. 1988). Oxygen is partitioned among anhydrous silicates, metal oxides, refractory organics, $CO$, $CO_{2}$ and $H_{2}O$ (Equations 2 and 3, below). To represent rock, we use a generic composition consisting of components $SiO_2 + MgO + Al_2O_3 + CaO + FeS + Ni$, with any remaining iron assumed to be present either as $Fe$ (in metallic alloy), $FeO$ (in silicates) or $FeO_{1.33}$ (equivalent to $Fe_3O_4$, magnetite). In our model, we consider anhydrous silicates ($\rho_{r}$$ \sim$ 3360 kg $m^{-3}$, Wong et al. 2008); metal sulfide+oxide phases ($\rho_{met}$$ \sim$ 4800 kg $m^{-3}$, Wong et al., 2008) or metal sulfide+native metal phases ($\rho_{met}$$ \sim$ 6300 kg $m^{-3}$, based on the cosmochemical abundances of Fe, S, Ni in Table 1). Except in the case of $FeO_{1.33}$, we consider these mixtures as primordial materials that formed KBOs to calculate the uncompressed densities. While metallic iron oxidation by water vapor in the protosolar nebula would not be efficient due to sluggish kinetics (Fegley 2000), we adopt $FeO_{1.33}$ as an end-member case because Pluto’s and Charon’s deep interiors may have experienced aqueous alteration during differentiation, and iron metal could have been oxidized to magnetite (McKinnon et al. 2017). Here, silicate minerals in the interiors of Pluto/Charon today may be hydrated (McKinnon et al. 2017); however, phyllosilicates can be considered as a combination of anhydrous minerals and structural water in hydrated silicate matrices, resulting in a similar overall density for the mixture (Waite et al. 2017). It is therefore appropriate to use the anhydrous equivalent to estimate the water budget and the overall density of the Pluto-Charon system. Following the method described in Johnson et al. 2012, we assume that water ice ($\rho_{w}$ $\sim$ 940 kg $m^{-3}$, Sloan \& Koh 2007) and $CO_2$ ice ($\rho_{CO_2}$ $\sim$ 1680 kg $m^{-3}$, Fard \& Smith 2024) are the primary condensed volatile ices in the outer protosolar nebula. Some $CO$ could have been incorporated into KBOs by being trapped in a water ice matrix such as clathrate hydrates (in a cool nebula) or directly condensed as $CO$ ice (in a cold nebula). For both cases, we set the amount of $CO$ incorporated into KBOs to be $\sim$ 3 \% by mole relative to water ice, as typically observed in cometary comae (Harrington Pinto et al. 2022), and the rest of $CO$ is assumed to be in the gaseous form in the protoplanetary disk (Prinn \& Fegley 1989; Krijt et al. 2020). We note that this probably represents an endmember scenario, as $CO$ ice does not appear to be abundant on the surface of Pluto, Sedna, Gonggong, Quaoar, Eris, Makemake (Glein \& Waite 2018; Emery et al. 2024; Grundy et al. 2024). Since clathrates have similar densities as water ice (Sloan \& Koh 2007), we assume that only the latter scenario (where $CO$ is present as a pure ice) would affect the overall uncompressed densities of KBOs. The contribution of $CO$ ice with $\rho_{CO}$  $\sim$ 880 kg $m^{-3}$ (Luna et al. 2022) for the cold nebula scenario is considered in addition to the above components. \\

Equations [1], [2], and [3] shown below can be solved together, which gives the atomic abundances of different elements partitioned among these material components. Finally, the resulting uncompressed density of KBOs can be calculated based on the last equation [4]. In the set of equations below, we introduce two observational parameters, $r_{org}$: fraction of protosolar C in refractory organics $r_{org} = n_{C-org} /n_C$; and $r_{CO_2}$: the relative molar abundance of $CO_2$ to water in comets $\sim$ 12 \%, including Jupiter-family comets and Oort cloud comets (Harrington Pinto et al. 2022). Total $CO$ in the protosolar nebula, including gaseous $CO$, $CO$ trapped in water ice and condensed $CO$ ice, is designated as $n_{CO}$. \\

The mass balance equations are:

\textbf{[1]}: $n_C = n_{CO} + n_{C-org} + n_{CO_2} =  n_{CO} + n_{C-org} + r_{CO_2}n_{H_2O}$

\textbf{[2a]}: $n_O = 2 n_{\text{Si[SiO}_2]} + 1.5 n_{\text{Al[Al}_2\text{O}_3]} + n_{\text{Ca[CaO]}} + n_{\text{Mg[MgO]}} + n_{\text{Fe[FeO]}} + n_{\text{O[org]}} + n_{\text{O[CO]}} + n_{\text{O[CO}_2]} + n_{\text{O[H}_2\text{O]}}$

\textbf{[2b]}: $n_O = 2 n_{Si[SiO_{2}]} + 1.5 n_{Al[Al_2O_3]} + n_{Ca[CaO]} + n_{Mg[MgO]} + 1.33 n_{Fe[Fe_3O_4]} + n_{O[org]} + n_{O[CO]} + n_{O[CO_2]} + n_{O[H_2O]}$

\textbf{[3]}: $n_{O[CO_{2}]} = 2n_{CO_2} = 2r_{CO_2}n_{H_2O}$; $n_{O[CO]} = n_{CO}$;  $n_{O[org]} = f_{O[org]}r_{org}n_{C}$, where $f_{O[org]}$ is the O/C ratio in refractory organics ($\sim$ 0.3 for comet 67P/C-G).\\

The mixture's density can be calculated via:

\textbf{[4]}: $\rho = (\frac{f_r}{\rho_r} + \frac{f_{met}}{\rho_{met}} + \frac{f_{org}}{\rho_{org}} + \frac{f_{ices}}{\rho_{ices}})^{-1}$, in which $f_i$ is the mass of each material component over the total mass, and $f_i = \frac{\mu_in_i}{\Sigma\mu_in_i}$ ($\mu_i$ is the molecular mas of each component). Here, $\frac{f_{ices}}{\rho_{ices}}$ is a general term that refers to ices that include water ice, $CO_2$ ice, and $CO$ ice. The density of the ice mixture is calculated using an equation of the same form as Equation 4. \\

\begin{figure}[ht!]
\plotone{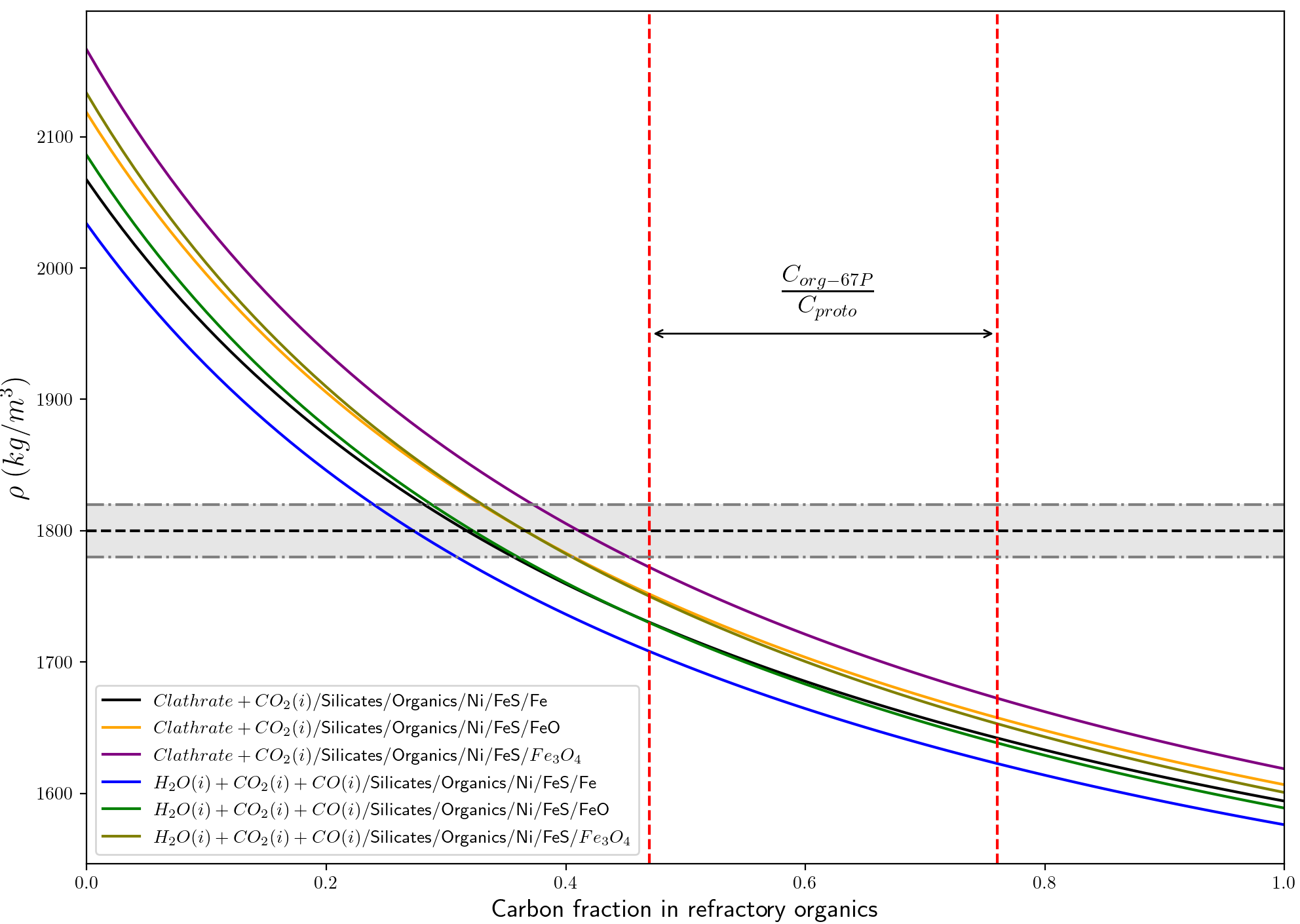}
\caption{The predicted uncompressed density of the Pluto-Charon system (colored curves) as the fraction of the protosolar C inventory contained in refractory organics increases from 0\% to 100\%, assuming the canonical protosolar abundances in Lodders 2021 (L21, see Table 1). The horizontal black dashed line represents the average uncompressed density of the system $\sim$ 1800 kg  $m^{-3}$ (Barr \& Schwamb 2016; McKinnon et al. 2017), taken to be the most representative value we have for the bulk composition of large KBOs. The gray area shows the 3$\sigma$ uncertainty ($\pm$ 20 kg   $m^{-3}$, Brozović et al. 2015, McKinnon et al. 2017, Nimmo et al. 2017).}  
\label{fig:general}
\end{figure}

Our results show a similar trend compared to the previous study by Wong et al. 2008 that the densities of KBOs should decrease with an increase in refractory organics. This trend arises because the oxygen-to-carbon ratio in organics (O/C $\sim$ 0.3, Bardyn et al. 2017) is less than that in $CO$ or $CO_2$; therefore, a higher fraction of protosolar C in refractory organics reduces the availability of C for $CO$ and $CO_2$ formation. This would free up more O to form water ice and lower the uncompressed density. As noted above, aqueous alteration of metallic iron to magnetite (i.e., during the differentiation of Pluto/Charon) would take up some amount of oxygen in rock, therefore resulting in a slight increase in the density of the Pluto-Charon system. Nonetheless, even when taking into account the 3$\sigma$ uncertainty in the uncompressed density of the Pluto-Charon system (1800 $\pm$  20  kg $m^{-3}$) (Brozović et al. 2015, McKinnon et al. 2017, Nimmo et al. 2017), the effect is not significant to increase the predicted density up to the lower limit of the measured range. In the cold nebula scenario where $CO$ ice is incorporated, the overall density is slightly lower than in scenarios where $CO$ is trapped in water ice. This difference is due to the lower density of $CO$ ice ($\sim$ 880 kg $m^{-3}$) compared with that of water ice ($\sim$ 940 kg $m^{-3}$). Surface observations of Pluto, however, suggest much less $CO$ ice ($CO/H_2O$ $\sim$ 4 $\times 10^{-6}$, Glein \& Waite 2018) than the amount considered in this calculation ($\sim$ 3\% relative to water, analogous to comets), so this is most likely an endmember scenario. In summary, we suggest that based on current values of the protosolar abundances, the predicted densities would be insufficient to explain the uncompressed density of the Pluto-Charon system ($\sim$ 1800 kg $m^{-3}$) (McKinnon et al. 2017, 2021). 

\section{Revised solar abundances: New constraints from solar neutrinos and solar wind with implications to KBOs and CI chondrites}\label{sec:new data}
\subsection{Data from the Sun}

In Section 1, we discussed the ongoing challenge in reconciling photospheric heavy element abundances with helioseismology constraints. Consequently, additional independent measurements are critical to test these models.  Solar neutrinos from the carbon-nitrogen-oxygen cycle were predicted to be generated by the seminal work of Hans Bethe (Bethe 1939) and were recently detected by the Borexino neutrino observatory with a significance of $\sim$ 7$\sigma$ (Agostini et al. 2020a). This opens a new door to constrain the solar abundances of C, N through neutrino flux measurements. These low-energy neutrino fluxes ($\phi(i)$) from the CNO cycle in the Sun’s interior are sensitive to the Sun’s temperature profile, which is controlled by the proton-proton (p-p) chains (Haxton \& Serenelli 2008). Consequently, if $\phi(i_1) \sim T^{(x1)}$ and $\phi(i_2) \sim T^{(x2)}$, one can form the weighted ratio $\phi(i_1)/\phi(i_2)^{x1/x2}$ that is nearly independent of temperature (Serenelli et al. 2013).  Thus, this analysis process breaks the degeneracy between opacity/composition and its variations that affect the interior’s temperature profile (Agostini et al. 2020b).  The flux of the ${}^{8}_{}B$ neutrino is highly sensitive to the solar core’s temperature ($\phi({}^{8}_{}B) \sim T^{24}$, Villante \& Serenelli 2021); thus, it has been used to probe the Sun’s interior temperature (Appel et al. 2022). Because the p-p chains control the temperature profile, the weighted ratio between ${}^{15}_{}O$ and ${}^{8}_{}B$ neutrino fluxes retains a linear dependence on the total surface abundance of C and N (after accounting for the efficiency of elemental diffusion) (Agostini et al. 2020b, Appel et al. 2022). As a result, this quantity can be directly compared with spectroscopic observations of the photosphere (Appel et al. 2022).\\ 

New findings suggest an increase in the total C+N abundance in the photosphere, with (C+N)/H $\sim (5.81_{-0.94}^{+1.22})\times 10^{-4}$, which display a $\sim 2\sigma$ tension with the low-metallicity solar model (Appel et al. 2022, Basilico et al. 2023). Since C is much more abundant than N in the Sun, most of that increase is likely attributed to the C abundance, with any reasonable changes in N imparting a secondary effect. The large error bars reflect uncertainties in the CNO neutrino fluxes – a challenging measurement given the physical nature of the neutrino. Despite this shortcoming, additional constraints from the water abundances in primitive solar system objects can be considered with appropriate caution to obtain further insights into the likely range of the C abundance. As demonstrated in the previous section (Section 2), the C abundance must be sufficiently higher than the currently adopted value to raise the density of large KBOs into the observational range; however, it must not be too high to generate KBOs with too little water. Here, we adopt an overall $\sigma$ value that matches the lower error bar of the solar neutrino measurements (to prevent a lack of water in KBOs); this gives (C+N)/H $\sim (5.81_{-0.94}^{+0.94})\times 10^{-4}$ for our subsequent calculations. As we demonstrate in Section 3.2, this constraint can be refined further by testing a range of plausible C/O ratios that would explain the water abundances in primitive outer solar system objects. \\

We combine this new result with a recent analysis of element abundances from Genesis solar wind data to estimate additional constraints on the protosolar abundances (Heber et al. 2021). To derive the abundances of N and O, it is necessary to consider fractionation from the photosphere to solar wind. Elemental fractionation is known to correlate with the first-ionization potential (FIP) of the elements; as a result, elements with low-FIP are overabundant in solar wind (Pilleri et al. 2015). As a first-order estimation, here we consider the elemental ratios of two elements with similar FIP values to be similar to their corresponding ratio in the photosphere. For N (FIP $\sim$ 14.53 eV, Heber et al. 2021), we consider Kr to be the closest analog with a FIP value $\sim$ 13.99 eV (Heber et al. 2021). In addition to having the most similar FIP to N, the solar Kr abundance can be reasonably well constrained based on the smoothness of the odd-even nuclide patterns and established abundances for neighboring nuclides from CI chondrites (Wiens et al. 1991). As a result, we can use the N/Kr ratio from the Genesis data with the Kr/H ratio from Lodders 2021 to derive the contemporary photospheric abundance of (N/H) $\sim(0.93 \pm 0.18) \times 10^{-4}$. While this is slightly higher than the value in Lodders 2021 $\sim (0.710 \pm 0.003) \times 10^{-4}$, it is consistent with a recent determination of N abundance from the 3D model by Magg et al. 2022 who estimated (N/H) $\sim (0.95 \pm 0.18) \times 10^{-4}$. Regardless, the N abundance has a minor impact on the overall metallicity. Combining this N/H with the (C+N)/H ratio from the solar neutrino data gives a present-day photospheric value for (C/H) $\sim (4.88 \pm 0.96) \times 10^{-4}$.\\ 

The next step is to constrain the oxygen abundance. Based on solar wind data from the Genesis mission, the fractionation model by Laming et al. 2017 (JML+17) suggests an increase in O/H compared with commonly adopted values. However, the Genesis data used in JML+2017 have been revised toward a decrease in both H and O fluences (Heber et al. 2021). Since H is much more abundant than O, this leads to an increase in O/H (the data point at HV+21, Fig. 2). Since O and H have similar FIPs ($\sim$ 13.6 eV, Heber et al. 2021), the fractionation between them from the photosphere to solar wind can be assumed to be minimal. To be more conservative given our assumption of a lack of O/H fractionation, here, we adopt a mid-range value for O that would satisfy both the previous analysis of Lamming et al. 2017 and the revised Genesis data, which gives a present-day O/H $\sim (6.68 \pm 0.49) \times 10^{-4}$. This range also overlaps with the more recent O values determined from spectroscopy by Magg et al. 2022 (Fig. 2). We recognize that our approach in constraining the O abundance is simplistic, which is why it is tested against independent data from KBOs and CI chondrites (see Section 3.2). Future solar neutrino experiments such as the Sudbury Neutrino Observatory + (The SNO+ collaboration 2021) and the Jiangmen Underground Neutrino Observatory – JUNO (Conroy 2024), where the latter is set to commence operations at the end of 2024, will enable further testing. To derive protosolar abundances from present-day photospheric values, we follow the approach described in Lodders 2020, 2021 to account for the settling effect of heavy elements. We converted photospheric abundances to protosolar values using a constant settling factor of $\sim$ 23\% for all heavy elements, as adopted in Lodders 2020, 2021. \\

\begin{figure}[ht!]
\plotone{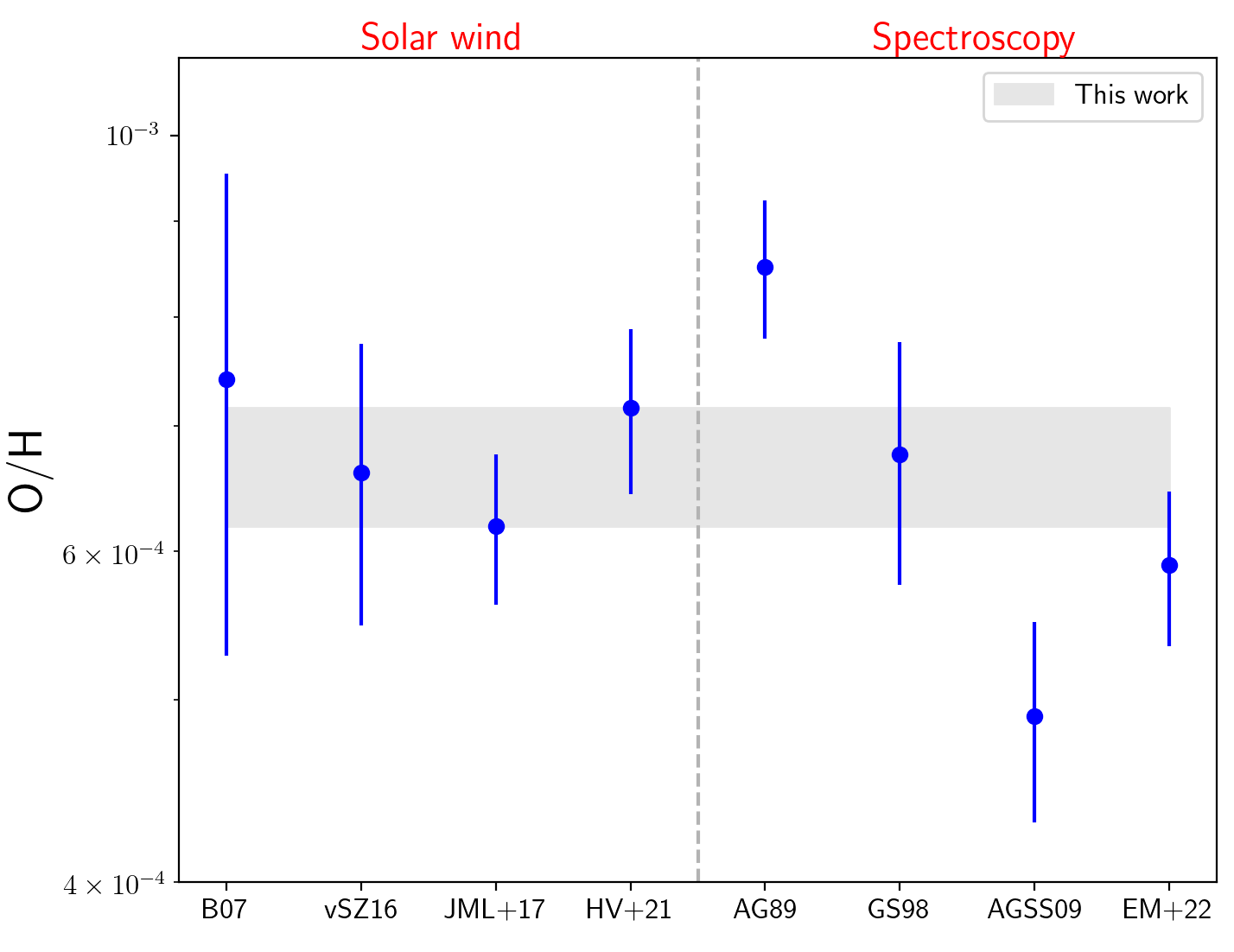}
\caption{Estimates of the present-day photospheric oxygen abundance from solar wind data (B07: Bochsler 2007, vSZ16: von Steiger \& Zurbuchen 2016, JML+ 2017: Lamming et al. 2017, HV+21: Genesis data from Heber et al. 2021) compared with spectroscopic determinations (AG89: Anders \& Grevesse 1989, GS98: Grevesse \& Sauval 1998, AGSS09: Asplund et al. 2009, EM+22: Magg et al. 2022). All uncertainties are reported here as $1\sigma$. The gray band represents the adopted range in this work. While past fractionation models based on in-situ mass spectrometry data in the Ulysses mission (vSZ16) and the Apollo lunar foil experiment (B07) reported estimates with large error bars, returned samples from Genesis allow more precise estimates with multiple concordances (JML+ 2017). The value in Lodders 2003 is the same as that in AGSS09, therefore it is not shown in this figure.}  
\label{fig:general}
\end{figure}

While estimates of the solar abundances have varied over time, spectroscopic measurements have shown an upward revision of the protosolar C/O ratio from 0.43 $\pm$ 0.05 in AG89 to 0.59 $\pm$ 0.08 in AAG21 and 0.62 $\pm$ 0.09 in EM+22 (Fig. 3). As noted in AAG21, the Sun has a higher C/O ratio than most solar twins regardless of age, which could be a result of the Sun having migrated rapidly outward since birth from a denser, more heavy element-rich region in the Milky Way (Nieva \& Przybilla 2012). In our subsequent calculations, we will adopt the average values of the C and O abundances in Table 2 (thus a C/O $\sim$ 0.73) as our nominal case and explore its implications for the compositions of KBOs and CI chondrites. We will also test a range of C/O ratios that could satisfy the observational densities from large KBOs (Appendix A). This ratio should be sufficiently above the currently adopted value to raise the density of KBOs to fall within the observed range; however, it must not be too high to produce KBOs with too little water. As we demonstrate in Figs. A1 and A2, the solar C/O is most likely within the range of 0.73 $\pm$ 0.10. The upper limit of our proposed C/O ratio is also consistent with the previous work by Pekmezci et al. 2019, who found that a negligible amount of water would exist in planetesimals under a C/O of 0.81, assuming that the entire protosolar C inventory is in $CO$. Under that assumption and given a C/O of 0.83, our calculation also arrives at a similar result. If some refractory organics contribute to the total protosolar C budget, this will allow the presence of water, though the amount is still insufficient to explain Pluto-Charon's density (Fig. A2). \\

\begin{figure}[ht!]
\plotone{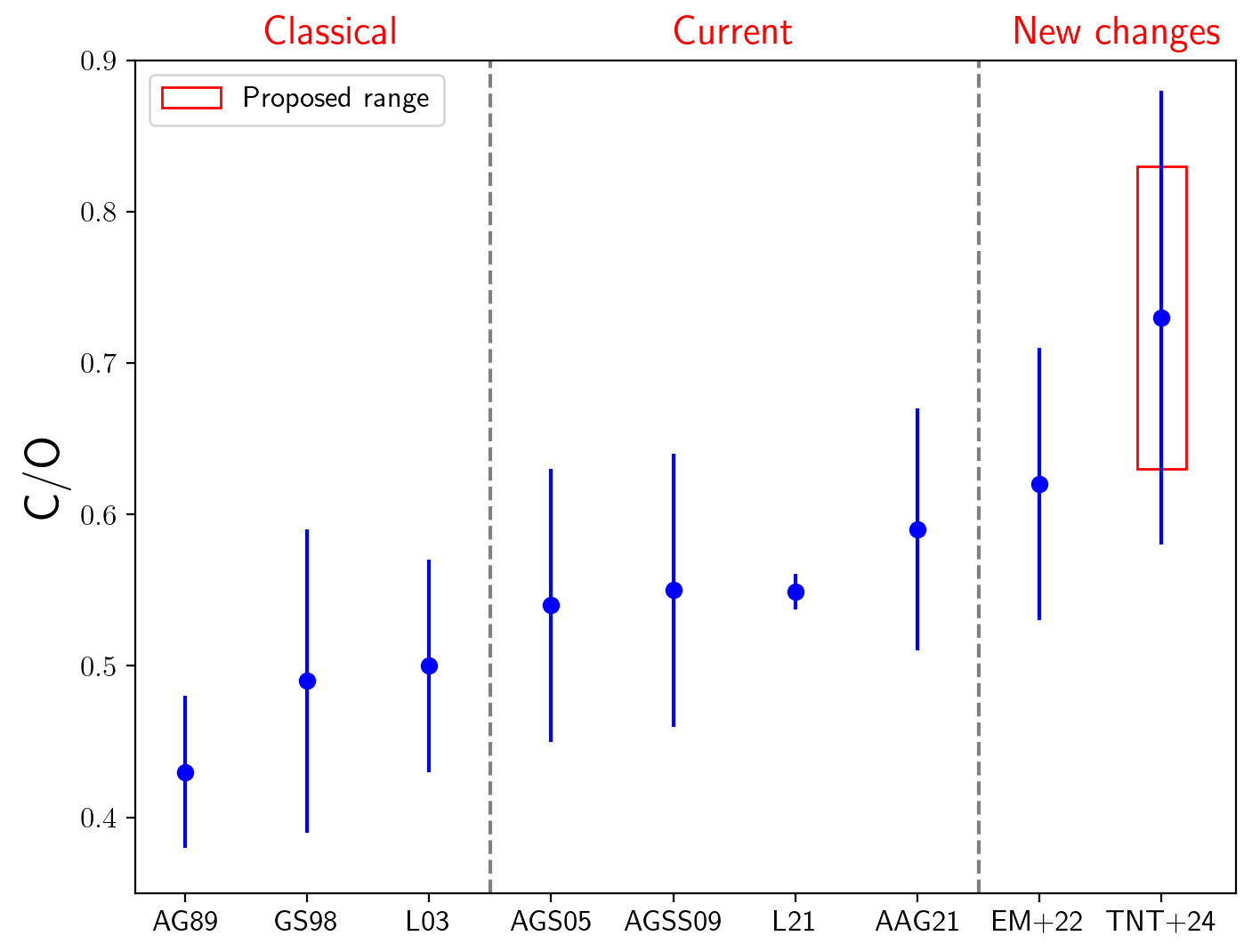}
\caption{Solar C/O ratio (0.73 $\pm$ 0.10) from this work (red box) compared with estimates in the literature through time. All uncertainties are reported here as 1$\sigma$. AG89: Anders \& Grevesse 1989, GS98: Grevesse \& Sauval 1998, L03: Lodders 2003, AGS05: Asplund et al. 2005, AGSS09: Asplund et al. 2009, L21: Lodders 2021, AAG21: Asplund et al. 2021, EM+22: Magg et al. 2022, TNT+24: this work.}  
\label{fig:general}
\end{figure}

\begin{figure}[ht!]
\plotone{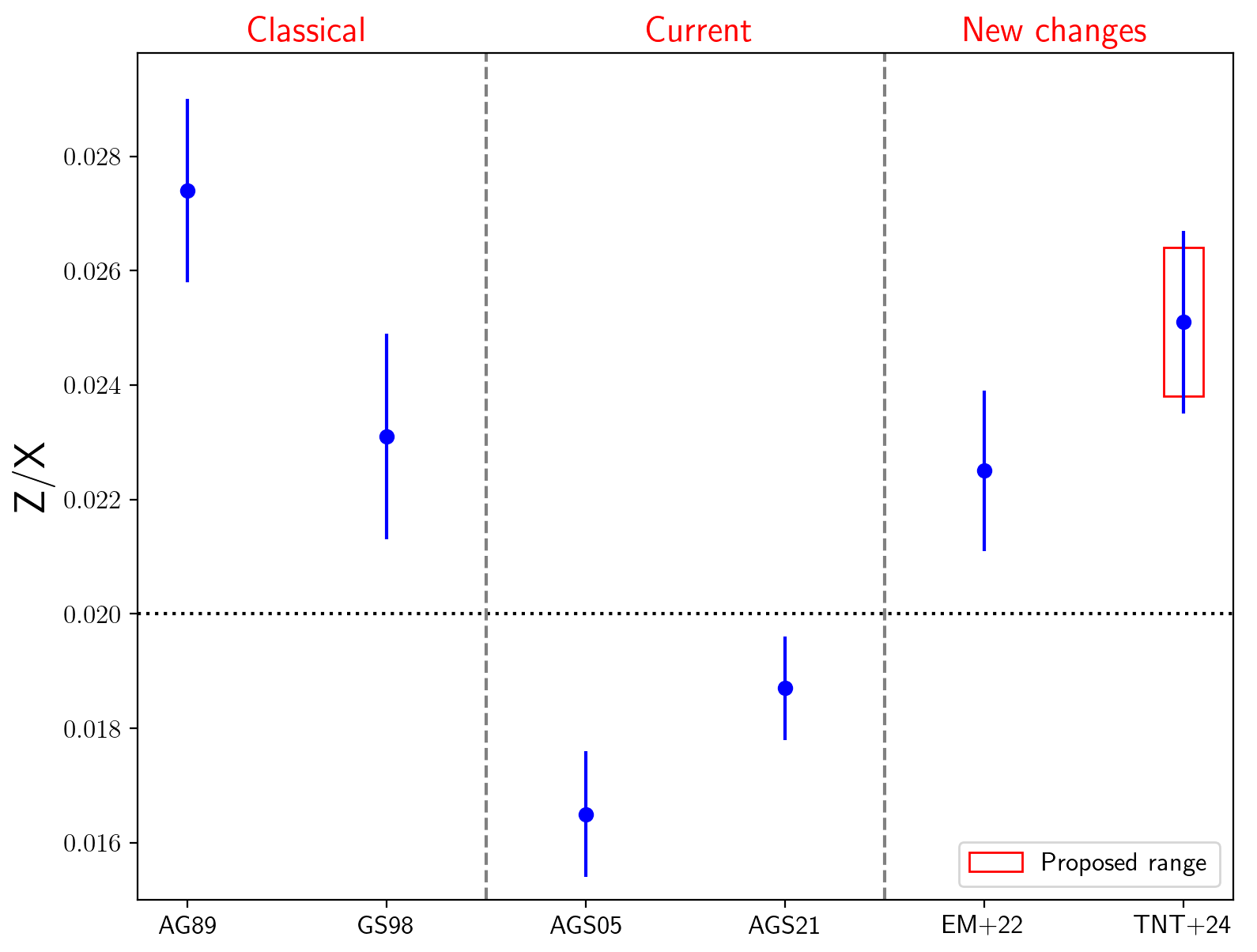}
\caption{Mass ratio of heavy elements to hydrogen in this study (Z/X = 0.0251 $\pm$ 0.0013, TNT+24) compared to previous works (AG89: Anders \& Grevesse 1989, GS98: Grevesse \& Sauval 1998, AGS05: Asplund et al. 2005, AAG21: Asplund et al. 2021, EM+22: Magg et al. 2022). The classical models here refer to the 1D, local thermodynamic equilibrium (LTE) radiative transfer models (Anders \& Grevesse 1989 - AG89; Grevesse \& Sauval 1998 - GS98). Despite their simplicity, these agreed well with constraints inferred from helioseismology such as the base of the convective zone ($R_{CZ}$), the sound speed and density profiles, and the He abundance in the top surface ($Y_S$) (Basu \& Antia 2008). Subsequent 3D models (AGS05, AGSS09, and AGS21) have decreased the abundances of several heavy elements, which led to disagreements with observations from helioseismology. AGSS09 reported $Z/X$ of 0.0181 and in Lodders 2003, this value is 0.0177. Both are within the range of values in the low-metallicity models by AGS05 and AGS21, therefore they are not shown here.}  
\label{fig:general}
\end{figure}

While C, N and O abundances went down starting with AGS05, the recent model by EM+22 revised those values upward. In the current work, the abundance of O is within the range of values in the classical high-metallicity model by GS98, and the current $Z/X$ is coincidentally consistent with values from AG89 and GS98 (Fig. 4). Our relatively high $Z/X$ may lead to better agreement with constraints from helioseismology than the current low-metallicity models (AGS05, AGSS09, AGS21). We note that the current $Z/X$ depends on the abundance of neon, and a recent analysis using updated atomic data has revised this value upward since GS98 (Young 2018). As neon does not appear in photospheric spectra and is largely lost in CI chondrites, it has been typically determined from the radiation originating in the high temperature regions of the Sun (i.e., corona, transition region, flares) (Young 2018) or solar wind (Huss et al. 2020). We followed the approach in Lodders 2021 to estimate the Ne abundance based on the Ne/O ratio in the transition region of the quiet Sun (0.24 $\pm$ 0.05) from Young 2018 and the oxygen abundance recommended here. The Ne/O ratio has also been revised upward from the adopted value $\sim$ 0.178 $\pm$ 0.035 in GS98 to 0.24 $\pm$ 0.05 (Young 2018), which may partly explain the higher value of our $Z/X$ compared with GS98. \\

We also note that not only the total heavy element abundances but also the ratios of these elements influence constraints from helioseismology, such as the depth of the convective zone and the speed of sound. If the values we recommended here could in future work lead to a better agreement with helioseismic constraints, then the mass fractions of hydrogen, helium, and heavy elements ($X$, $Y$, and $Z$, respectively) can be determined using an $X$ or $Y$ determination from helioseismology, given our $Z/X$ value and $X + Y + Z = 1$. Previous solar models found that the estimated $X$ for the Sun (0.7389 $\pm$ 0.0034) appears to be fairly independent of $Z$ (Basu \& Antia 2004, 2008), so we use this value of $X$ to calculate $Y$ and $Z$. Given $Z/X$ = 0.0251 $\pm$ 0.0013 (Table 2), the corresponding values for $Y$ and $Z$ are: $Y = 0.2426 \pm 0.0035$, $Z = 0.0185 \pm 0.0010$. The helium mass fraction $Y$ from our model is similar to the helium abundance from solar models using GS98 heavy element abundances (0.2425 $\pm$ 0.0042) (Basu \& Antia 2008). Indeed, both values overlap with the surface helium abundance determination from helioseismology (0.2485 $\pm$ 0.0034, Basu \& Antia 2004, 2008). In order to estimate the corresponding protosolar values, we adopt the protosolar helium mass fraction determination from solar models by Serenelli \& Basu 2010, $Y_{proto} = 0.2780 \pm 0.0060$. Given $(Z/X)_{proto} = 0.0308 \pm 0.0016$ (Table 2), the corresponding values for $X_{proto}$ and $Z_{proto}$ are: $X_{proto} = 0.7004 \pm 0.0059$, $Z_{proto} = 0.0216 \pm 0.0011$. \\

{

\begin{tabular}{ |c|c|c|c|c|}
 \hline

\ {Abundances} & \multicolumn{2}{|c|}{Present} & \multicolumn{2}{|c|}{Protosolar}\\
\hline

\ {Element} & N(E) & $\pm\sigma$&N(E) & $\pm\sigma$\\
 \hline

H   & $3.09\times 10^{10}$ & & $2.52\times 10^{10}$ & \\

He  & $2.54\times 10^{9}$  & $3.8\times 10^{7}$ & $2.50\times 10^{9}$ & $5.8\times 10^{7}$ \\

C   & $1.51\times 10^{7}$  & $1.8\times 10^{6}$ & $1.51\times 10^{7}$ & $1.8\times 10^{6}$ \\

N   & $2.87\times 10^{6}$  & $6.0\times 10^{5}$ & $2.87\times 10^{6}$ & $6.0\times 10^{5}$ \\

O   & $2.07\times 10^{7}$  & $1.5\times 10^{6}$ & $2.07\times 10^{7}$ & $1.5\times 10^{6}$ \\

Mg  & $1.03\times 10^{6}$  & $4.4\times 10^{4}$ & $1.03\times 10^{6}$ & $4.4\times 10^{4}$ \\

Si  & $1.00\times 10^{6}$  & $3.4\times 10^{4}$ & $1.00\times 10^{6}$ & $3.4\times 10^{4}$ \\

Al  & $81820$              & $6110$            & $81820$             & $6110$             \\

Ca  & $57239$              & $4500$            & $57234$             & $4500$             \\

Ni  & $48670$              & $2940$            & $48670$             & $2940$             \\

Fe  & $8.72\times 10^{5}$  & $3.8\times 10^{4}$ & $8.72\times 10^{5}$ & $3.8\times 10^{4}$ \\

S   & $4.37\times 10^{5}$  & $2.6\times 10^{4}$ & $4.37\times 10^{5}$ & $2.6\times 10^{4}$ \\
 \hline
\end{tabular}

\textbf{Table 2}. Present (left) and protosolar abundances (right) estimated in this work from solar neutrinos and solar wind data for C, N, and O. The error bar for C is determined by the uncertainties reported from the solar neutrino and solar wind measurements, with additional constraints from a range of C/O ratios that can satisfy the observational densities of large KBOs (Appendix A). As heavy elements have gravitationally settled from the outer convective zone into the Sun’s interior over the lifetime of the Sun, the protosolar abundances relative to $10^{12}$ hydrogen atoms ($N(E)/N_{H-proto}$) need to be corrected from the present abundances relative to $10^{12}$ hydrogen atoms ($N(E)/N_{H-present}$) (Palme et al. 2014; Lodders 2020, 2021). The atomic ratio between He and H is given by: $\frac{N(He)}{N(H)} = \frac{Y}{4X}$ where the ``4'' is shorthand for the ratio of the atomic weights of He and H. Here, the present-day mass fractions of hydrogen, helium, and heavy elements ($X$, $Y$, and $Z$, respectively) are: $X = 0.7389 \pm 0.0034$, $Y = 0.2426 \pm 0.0035$, $Z = 0.0185 \pm 0.0010$ and the corresponding protosolar values are: $X_{proto} = 0.7004 \pm 0.0059$, $Y_{proto} = 0.2780 \pm 0.0060$, $Z_{proto} = 0.0216 \pm 0.0011$.

\subsection{Independent tests based on KBOs and CI chondrites\label{subsec:test}}

We repeat the calculation from Section 2 to see how much the predicted density of Pluto-Charon changes with our new solar C/O $\sim$ 0.73. This is shown in Fig. 5. A notable difference compared with Fig. 1 is a reduction in the fraction of protosolar C as refractory organics in the KBO-forming region ($20-30$ AU, McKinnon et al. 2021), which is attributed to the overall increase in the protosolar C abundance. The overall increase of the C/O ratio also results in less O available to form water, thereby favoring the accretion of refractory components over water ice in KBOs. Both effects lead to an increase in the overall densities compared to results in Fig. 1. \\

\begin{figure}[ht!]
\plotone{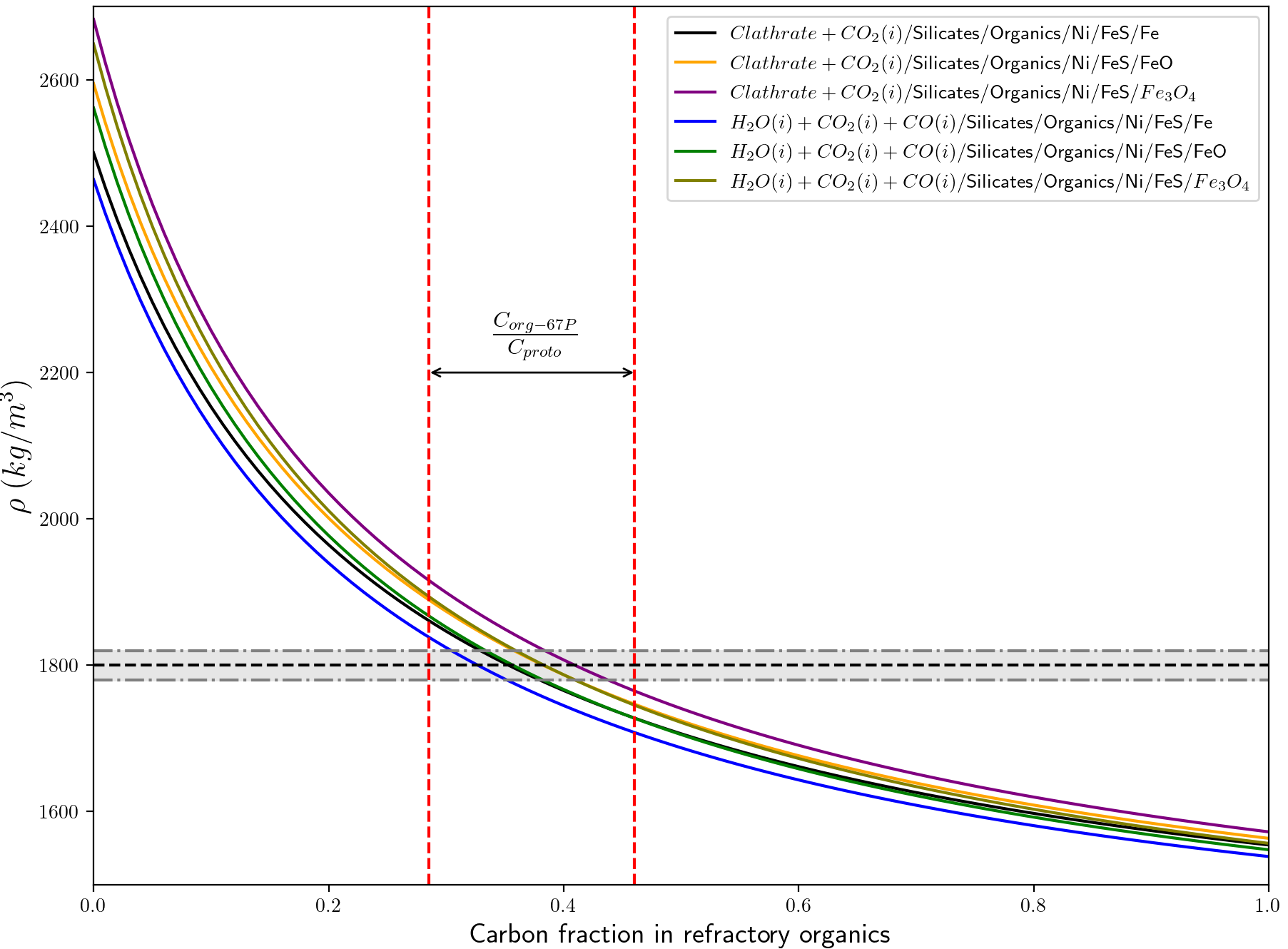}
\caption{The predicted uncompressed density of the Pluto-Charon system (colored curves) as the fraction of the protosolar C inventory contained in refractory organics increases from 0\% to 100\% under the new protosolar abundances (Table 2). The horizontal black dashed line represents the average uncompressed density of the system $\sim$ 1800 kg  $m^{-3}$ (Barr \& Schwamb 2016; McKinnon et al. 2017), taken to be the most representative value we have for the bulk composition of large KBOs. The gray area shows the 3$\sigma$ uncertainty ($\pm$ 20 kg   $m^{-3}$, Brozović et al. 2015, McKinnon et al. 2017, Nimmo et al. 2017).}  
\label{fig:general}
\end{figure}

Given the updated protosolar abundances and a range for the abundance of refractory organics constrained by comet 67P/C-G data, the predicted densities are found to be consistent with the 3$\sigma$ uncompressed density of the Pluto-Charon system. Within the given range of refractory organics inventory in protosolar C, our model can also predict KBOs that are more rock-rich than the Pluto-Charon system, up to 1900 kg $m^{-3}$ $-$ the uncompressed density of Triton today (Wong et al. 2008). Given the origin and evolution of Triton, which may include volatiles that were lost post-capture as discussed in Section 1, Triton’s case can be considered as an upper limit for the uncompressed density of large KBOs. These findings do not depend on the adopted value of the settling factor. As discussed in Section 3.1, heavy element settling in the Sun only affects the protosolar abundances relative to hydrogen; the relative ratios among the heavy elements are approximately constant. Therefore, on the cosmochemical scale (relative to silicon), the contemporary abundances of heavy elements stay the same as the protosolar abundances (Lodders 2020, 2021). \\

Comets are also ice-rich objects, so in principle, they could provide a related test of the solar oxygen abundance if their total O/Si ratios can be measured. However, the bulk density cannot serve as a reliable indicator since comets contain substantial porosity (Pätzold et al. 2019). Nevertheless, the refractory-to-ice mass ratio (R/I) depends on the elemental abundance of oxygen, and it can be constrained using observational data. Under this updated set of solar abundances data, we predict the R/I ratio within comet nuclei to be approximately $\sim$ $1.9-2.5$ (Fig. B1). Our mass balance calculation can be seen as a refinement of the model of Greenberg 1998. While comet 67P/C-G is an evolved comet and it is challenging to derive the R/I ratio for the nucleus, our predicted value falls within the range of plausible values  ($\sim$ $1-6$) derived by various investigations on Rosetta (Choukroun et al. 2020). The R/I ratio implied by our solar C/O ratio could be further tested by a future comet sample return mission such as CAESAR (Hayes et al. 2020).\\

The uncompressed densities of the the major icy satellites of Jupiter and Saturn (Ganymede, Callisto, Titan), with values near $\sim$ 1500 kg $m^{-3}$ (Wong et al. 2008), correspond to more ice-rich compositions than in the large KBOs. As discussed in earlier work (e.g., Simonelli et al. 1989, Wong et al. 2008), the range in density among icy satellites requires different carbon chemistry and/or significant fractionation of ice and rock in subnebulae around giant planets, where these regular satellites are thought to have formed (Lunine \& Stevenson 1982; Canup \& Ward 2002). In principle, a more reducing $CH_4$-rich (or very organic-rich) subnebula could produce more water-rich materials as has been suggested by equilibrium chemistry models under conditions in the subnebulae of giant planets (i.e., Lunine \& Stevenson 1982; Prinn \& Fegley 1989). It has also has been hypothesized that the formation of giant planets led to a more oxygen- or water-rich subnebula than the protoplanetary disk, depending on the relative rates at which different types of planetesimals dissolved in the early envelopes of giant planets (Podolak et al. 1988; Simonelli et al. 1989; Mosqueira \& Estrada 2003). Alternatively, the outward diffusion and recondensation of water vapor from dehydrated chondritic materials in the Jovian subnebula would allow condensation of significant amounts of ice in the formation region of Ganymede and Callisto (Mousis et al. 2023). For the Saturnian icy satellites, there remains the possibility that they were formed as collisional products of primordial differentiated satellites earlier in Saturn’s history (Canup \& Ward 2006; Ćuk et al. 2016). As a result, significant redistribution of ice and rock might have occurred, and satellite compositions might reflect the resulting fragments from past collisions.\\  

We further tested the new solar abundances (Table 2) with compositional data from CI chondrites. Since they are thought to have formed in the outer solar system at the greatest heliocentric distance among chondrites (Kruijer et al. 2017; Desch et al. 2018), the inventory of primordial water in their parent bodies can be expected to reflect the protosolar oxygen abundance. In order to quantify the primordial amount of water ice accreted in the CI chondrite parent body, one must account for various sources as many elements like Fe, S, and P can be oxidized by water during aqueous alteration. Previous estimates considered phyllosilicates, carbonates, iron oxides, phosphates, and sulfates as secondary mineral products of aqueous alteration in CI chondrites (Alexander 2019). However, the indigenous nature of sulfates in chondrites has been questioned and a terrestrial origin from oxidation of sulfides has also been proposed (i.e., Gounelle \& Zolensky 2001, 2014}; Airieau et al. 2005). Recent missions such as Hayabusa-2 and OSIRIS-REx returned samples from asteroids Ryugu and Bennu, respectively, and analyses of these samples provide new insights into the composition of primitive solar system bodies. Analyses of Ryugu samples suggest a similar mineralogy to CI chondrites with abundant phyllosilicates, carbonates, iron sulfide, and magnetite; this indicates low temperature, reducing, alkaline aqueous alteration, rather than the oxidizing conditions required for sulfate formation (Yokoyama et al. 2023). Abundant carbonate and the bulk abundances and isotopic ratios of titanium and chromium suggest that Ryugu formed in the outer solar system, beyond the water and $CO_2$ snowlines (Nakamura et al. 2023; Yokoyama et al. 2023). A notable distinction from CI chondrites is Ryugu’s lack of sulfates and ferrihydrite (Yokoyama et al. 2023). Rather than sulfates, sulfur appears to be present in the form of sulfide in pyrrhotite (Yokoyama et al. 2023; Nakamura et al. 2023). This apparent absence, along with the offset in $\Delta{}^{17}_{}O$ between Ryugu and CI chondrites suggest that large abundances of sulfates in carbonaceous chondrites are due to terrestrial contamination (Yokoyama et al. 2023). Similar to Ryugu, early analyses of Bennu samples also indicate the presence of phyllosilicates, Fe-sulfides, magnetite, and carbonates with no evidence for sulfate (Lauretta et al. 2024, King et al. 2024). Hence, previous estimates of accreted water ice content in CI chondrites (e.g., Alexander 2019) based on the presence of sulfates may now be seen as too large. This finding highlights the importance of recent asteroid sample return missions in refining our knowledge of the initial compositions of primitive solar system bodies. \\

In light of the new data from Ryugu and Bennu, we revisited the previous calculations of Alexander, 2019 to more accurately constrain the total amount of excess O initially accreted as water ice ($O_{ex}$) in the CI parent body. Excess O is defined to be the amount of oxygen that is above what is needed to charge-balance metal cations in accreted anhydrous minerals. Here, we consider water (including interlayer $H_2O$, and structural $H_2O$ and $OH$ in phyllosilicates) (Yokoyama et al. 2023) and carbonates, iron oxide, and phosphate as products of aqueous alteration (Alexander 2019) to calculate $O_{ex}$. Although sulfur is also found in refractory organics in CI chondrites (Alexander et al. 2017) and Ryugu samples (Yamaguchi et al. 2023, Yoshimura et al. 2023, Takano et al. 2024),  its abundance in organics is significantly lower compared to that in iron sulfides (Alexander et al. 2017). Given that sulfur in refractory organics constitutes 1-3 atoms per 100 carbon atoms (Alexander et al. 2017) and considering the abundance of refractory organics in CI chondrites and Ryugu samples ($\sim$ 2-3 wt\%, Yokoyama et al. 2023), sulfur in organics could represent only $\sim$ 0.7-4 \% of the total sulfur. Therefore, for this calculation, we adopted the approach proposed by Alexander 2019, assuming that the majority of sulfur was initially accreted into chondrites as $FeS$, with the remaining iron predominantly in metallic form. We calculate the mass of water consumed ($m_w$) when Fe was fully oxidized to magnetite through the reaction $3Fe + 4H_2O \rightarrow Fe_3O_4 + 4H_2$. Here, $m_w = 1.33 \mu_w \left(\frac{m_{\text{Fe-bulk}}}{\mu_{Fe}} - \frac{m_S}{\mu_S}\right) \sim 39 \, \text{g} \, \text{kg}^{-1}$ of CI chondrites, given $m_{\text{Fe-bulk}} = 185 \, \text{g} \, \text{kg}^{-1}$ of CI, $m_S = 53.5 \, \text{g} \, \text{kg}^{-1}$ of CI (Table 1 in Alexander 2019). The equivalent amount of O in water consumed by iron oxidation is given in Table 3 by scaling the molar mass of water (18 g $mol^{-1}$) with that of oxygen (16 g $mol^{-1}$). This amount is much lower than Alexander’s 2019 estimate of 71.2 g $kg^{-1}$ of CI, which was computed by assuming that a large fraction of sulfur (\textgreater 91 \%) was in the form of sulfate, leading to most iron being oxidized to iron oxides.\\

Yet, the amount of reacted water should be treated as a lower limit for the amount of accreted water because it might be possible that the amount of water that was present exceeded the chemical capacity of the rock. Some unreacted water could have been trapped within the interior cracks and voids between mineral grains and might have been lost during the subsequent evolution of the parent body. While there are various factors that influence pore volume (i.e., degassing of other volatiles, space weathering, impact, etc.), we consider an endmember scenario in which the volume consisting of present-day cracks and voids after aqueous alteration was formerly filled with liquid water to derive an upper limit on the inventory of accreted water (Table 3). For Ryugu samples, this volume fraction is $\sim$ 12 $\pm$ 4\% (Tsuchiyama et al. 2024), which we consider the upper value (16\%) to derive an upper bound on the amount of water that was physically removed after aqueous alteration. It can be calculated as:\\

\textbf{[5]}: $\frac{m_w}{m_r} = \frac{\rho_{w-l}\phi}{\rho_{r}(1-\phi)}$\\

Here $\phi$ is the porosity in present-day cracks and voids, $\rho_{w-l} \sim$ 1000 kg $m^{-3}$ is the density of liquid water, $\rho_{r}$ is the dry grain density, taken to be $\sim$ 3500 kg $m^{-3}$ (Macke et al. 2011), $m_r$ is the mass of accreted anhydrous rock (g) per kg of CI $\sim$ 700 g $kg^{-1}$ (Table 1 in Alexander 2019). The exclusion of sulfates from our calculations, based on their lack of detection in Ryugu and Bennu samples, markedly decreases the estimated $O_{ex}$ (Table 3). This adjustment results in a roughly $\sim$ $30-50$ \% decrease in the previously estimated value of 259.2 g $kg^{-1}$ from Alexander 2019.\\

Alternatively, the initial amount of water ice accreted into CI chondrite parent bodies can be constrained based on the volume ratio between liquid water and rock just before the onset of aqueous alteration, assuming that all pore space prior to aqueous alteration was filled with water (Bouquet et al. 2021). While the rocky component today is aqueously altered, the porosity within the phyllosilicate matrix would reflect the volume expansion of anhydrous silicates to phyllosilicates; that volume was previously occupied by liquid water prior to the onset of aqueous alteration. An upper limit case can be considered by assuming that the total porosity within the phyllosilicate matrix and present-day cracks and voids was all filled with water. Measurements from Ryugu samples indicate a total porosity of (42 $\pm$ 8) \% (Nakamura et al. 2022, Tsuchiyama et al. 2024). For only one CI sample that was measured, the  total porosity is slightly lower ($\sim$ 34.9 $\pm$ 2.1 \%, Macke et al. 2011), but is consistent within the error bar of Ryugu samples measurements. Using the upper end of the error bar for the total porosity ($\sim$ 50 \%) and Equation 5, we derive an upper limit for the amount of water initially present prior to the onset of aqueous alteration. In this case, the equivalent O initially accreted as water ice is $\sim$ 178 g $kg^{-1}$ of CI. This value falls within the range of the total excess O derived by considering the total amount of water consumed during aqueous alteration and the amount of excess water lost post-alteration (Table 3). This consistency between two estimation methods typically used in the literature provides additional support for the results presented here. \\

{
\centering
\begin{tabular}{|c|c|}

\hline

\ {Reservoir of O} & {Abundances (g per kg of CI)}\\
\hline

O in carbonate   & $0 - 4.8$${}^{(a)}_{}$   \\
O in phyllosilicate $H_2O$ & $107.6 - 118.8$${}^{(a,b)}_{}$ \\
O in Fe oxide & 35.0 \\
O in phosphate & 1.3${}^{(a)}_{}$   \\
O in unreacted $H_2O$ & $0 - 31.1$ \\
Total excess O ($O_{ex}$) & $143.9 - 191.0$ \\

\hline
\end{tabular}

}

\textbf{Table 3}. Data used for calculating the total excess O initially accreted as water ice assuming that the rocky components of CI chondrites were anhydrous with all S in $FeS$, all remaining Fe in metal, and all P in phosphide. ${}^{(a)}_{}$Alexander 2019; ${}^{(b)}_{}$Yokoyama et al. 2023. Since carbonate formation may not require O derived from water (if $CO_2$ was primordial: e.g., $CaO + CO_2$ $\rightarrow$ $CaCO_3$), we adopt a range for carbonate from 0 to 4.8 g $kg^{-1}$ of CI, the latter value was reported in Alexander 2019. For phyllosilicate $H_2O$, more recent measurements of CI chondrites from Yokoyama et al. 2023 are considered; the datum used in Alexander 2019 is within that range. The average value for O in phosphate was reported in Alexander 2019. \\

\begin{figure}[b]
\plotone{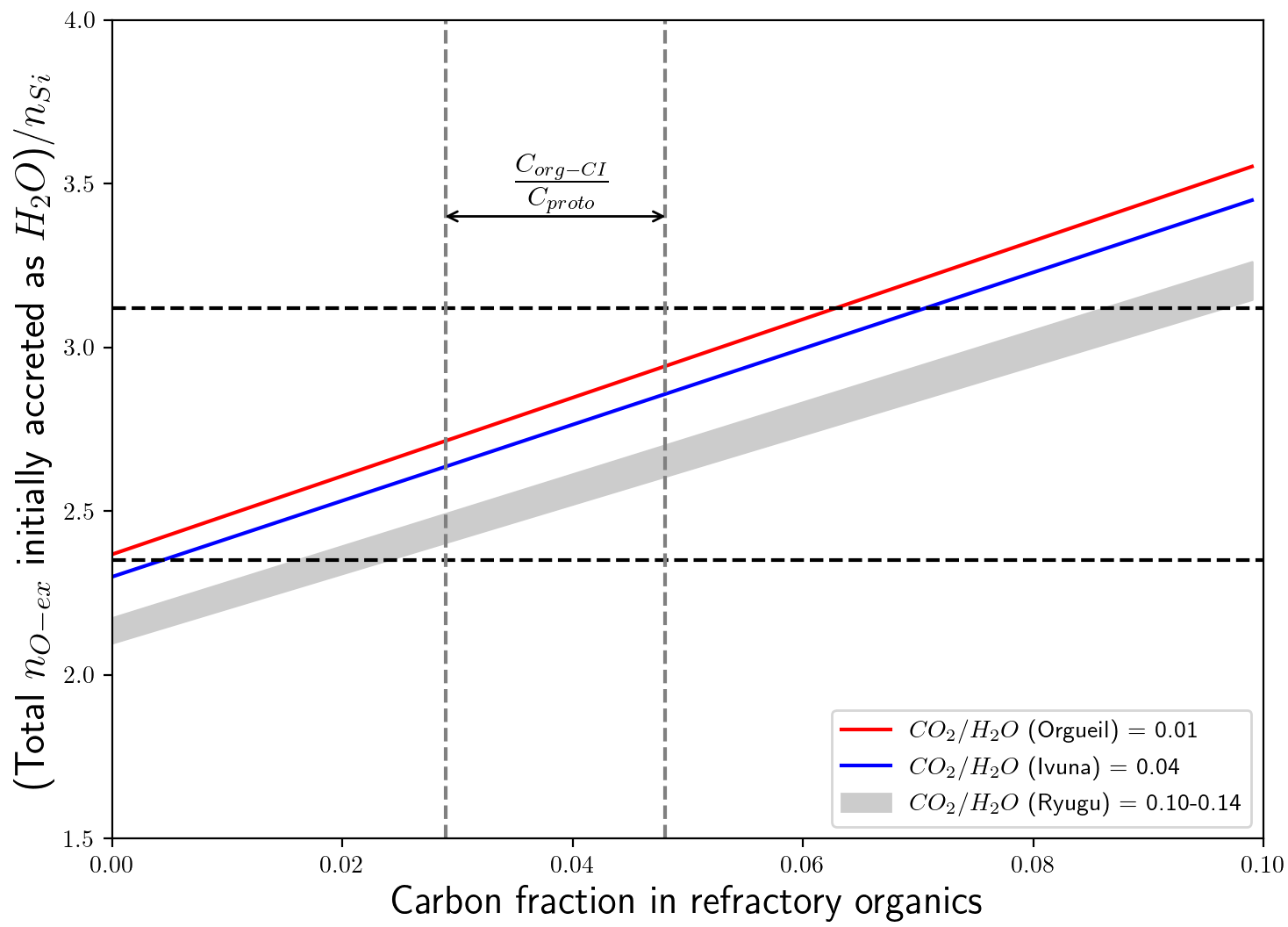}
\caption{The observationally constrained total water ice initially accreted into CI chondrite parent bodies (dashed black lines) ($n_{O_{ex}}$), including new constraints from Ryugu and Bennu samples, compared to the predicted range (colored lines) of water abundances in CI chondrites and Ryugu (gray band) for the revised solar abundances proposed in this work. Due to space weathering (Appendix C), the lower bound of the constrained water ice initially accreted into Ryugu may be lower than that of CI. An upper limit for the equivalent O initially accreted as water ice per silicon atom in Ryugu is estimated to be $\sim$ 2.9 based on the upper end value for the total porosity $\sim$ 50 \% (Nakamura et al. 2022, Tsuchiyama et al. 2024) and Equation 5. The predicted water content for Ryugu (gray band) falls within the observationally constrained total water ice initially accreted into CI chondrite parent bodies (dashed black lines), making it consistent with Ryugu's equivalent constraints. Note that the abundance of organic carbon in CI is about ten times smaller than that from comet 67P (which is a compositional model for Kuiper belt objects). This organic carbon abundance in Ryugu is similar to CI.}
\label{fig:general}
\end{figure}

We are now ready to see whether our solar abundances yield a primordial water inventory consistent with the range constrained by CI chondrite and Ryugu data (Table 3). We construct a mass balance model similar to our previous calculations for KBOs but modified to account for CI-specific compositions. Here, we consider a set of accreted materials including anhydrous silicates, native metals, refractory organics, and ices (including water ice and $CO_2$ ice, as CI chondrites show evidence of having formed beyond the water and $CO_2$ snowlines (Nakamura et al. 2023). Unlike $CO_2$, $CO$ is assumed to be a gas in the region of the protosolar nebula where CI chondrites formed (as suggested by their relatively low D/H ratios in water, e.g., Alexander et al. 2012, Piani et al. 2023), and therefore is not directly accreted by these chondrites in solid form. For refractory organics, we consider insoluble organic matter (IOM) found in CI chondrites, which has an average composition of $C_{100}H_{80}O_{20}N_4$ (Alexander et al. 2017). The carbon content within refractory organics in CI chondrites ranges between 2.0 wt.\% to 3.3 wt.\% ($20-33$ g $C_{org}$ per kg of CI, $m_{C-org}$), as reported in Alexander et al. 2007 and Yokoyama et al. 2023. With 107 g of silicon ($m_{Si}$)  per kg of CI chondrite (Alexander 2019), the abundance of refractory organic carbon ($C_{org-CI}$) on the cosmochemical scale (relative to silicon) can be calculated as: $C_{org-CI} = \frac{m_{C}\mu_{Si}}{m_{Si}\mu_{C}} $ where $\mu_{Si}$ and $\mu_{C}$  are the molar masses of Si (28 g $mol^{-1}$) and C (12 g $mol^{-1}$), respectively. In this calculation, we adopt a relative molar abundance of accreted $CO_2$ (ice) to water ice ($r_{CO_2}$) of 1-4 \% based on measured carbonate abundances in CI chondrites (Alexander et al. 2015). Ryugu samples are considered more pristine and free from terrestrial contamination than CI chondrites are (Yada et al. 2022, Nakamura et al. 2023). However, they also underwent modification due to space weathering (Nishiizumi et al. 2022, Yokoyama et al. 2023), as further discussed in Appendix C. Consequently, estimating the $CO_2/H_2O$ ratio based on the total hydrogen and carbonate content remains difficult (Alexander et al. 2015). Given that Ryugu formed beyond the $CO_2$ snowline and appears to contain more carbonates than CI chondrites (Yokoyama et al. 2023), we adopted a $CO_2/H_2O$ ratio range of 0.10-0.14, which is typically observed in comets (Harrington Pinto et al. 2022). This ratio may be refined through ongoing and future studies aimed at measuring the $CO_2/H_2O$ ratio in fluid inclusions (e.g., Tsuchiyama et al. 2021, Zolensky et al. 2022). We solve Equations 1-3 with these parameters  to predict from our model the initial amount of water ice in CI chondrite and Ryugu parent bodies relative to silicon ($n_{H_2O}$) in Equation 2; this value is equivalent to the number of moles expressed as $O_{ex}$ ($n_{O_{ex}}$).\\

Figure 6 shows the predicted abundance of water ice accreted into CI chondrite and Ryugu parent bodies based on the new set of solar abundances. A higher C/O ratio as proposed in this work leads to a decrease in water content in planetesimals in the region of the solar nebula where CI chondrites formed; this is consistent with the downward revision of the amount of water ice accreted into CI chondrites, given the most up-to-date constraints from Ryugu and Bennu samples. The revised solar abundances predict the compositions of outer solar system objects ranging from CI chondrites to Kuiper belt objects – representing a vast region of the outer solar system. This consistency further supports the new data from solar neutrinos and solar wind measurements. The present interpretation can be further tested with future data from ongoing analyses of Bennu samples, which early results suggest to be more organic-rich  than Ryugu or CI chondrites (Lauretta et al. 2024). Given a bulk C abundance of $\sim$ 4.6 wt.\% in which 90\% of the C is in refractory organics (Lauretta et al. 2024), we predict that the total amount of water ice initially accreted into Bennu’s building blocks is $\sim$ $210$ g $kg^{-1}$. Due to the higher abundance of refractory organics, we suggest that Bennu samples should be more water-rich than CI chondrites and/or contain higher porosity. \\

Lastly, it has not escaped our attention that the initial water/rock ratio, thus the water abundance itself, in principle can also be inferred by considering oxygen isotopic exchange during the parent body aqueous alteration process (Clayton \& Mayeda 1999; Brearley 2006; Fujiya 2018). Even though any extra water would have escaped, its abundance would have left its mark on the oxygen isotopic composition of the altered rock. This method, however, is subject to other complications, most notably the unknown oxygen isotope ratio in the initial water and the poorly defined oxygen isotope ratios in the anhydrous minerals prior to the onset of aqueous alteration. Thus, additional assumptions are required (Clayton \& Mayeda 1999). Although some anhydrous grains survived the alteration process, it is not clear that representative oxygen isotope values in CI chondrites can be well-defined due to significant grain-to-grain variations (Leshin et al. 1997, Clayton \& Mayeda 1999). Similar to previous findings in CI chondrites (Leshin et al. 1997), Ryugu's anhydrous minerals display a bimodal distribution in oxygen isotope ratios (Kawasaki et al. 2022), which underscores the difficulty. Given these complexities, it may be most useful to flip the problem and instead use our predicted water-to-rock mass ratio to estimate the primordial oxygen isotope ratios in water, which can be expressed as $\delta_{w}^{18} = 34.0-44.8$ {\textperthousand}  and $\delta_{w}^{17} = 17.7-23.4$ {\textperthousand} relative to the isotopic standard of Vienna Standard Mean Ocean Water (Appendix D). These predictions could potentially be tested by a future sample return mission from comet Hartley 2, as it has water with a D/H ratio similar to that of CI chondrites (Hartogh et al. 2011).\\

\section{Concluding remarks\label{subsec:conclusion}}
This work argues for refractory organics as one of the main C carriers in the outer protosolar nebula in the KBO-forming region. Incorporating their presence is not just a mere compositional detail, but indeed, they significantly influence the chemical makeup of the nebula and the building blocks that formed planetary bodies. By utilizing Rosetta COSIMA data for comet 67P/C-G, we show that under the current widely adopted solar abundances, refractory organics would make up a very large fraction of the protosolar C inventory in the KBO-forming region. As a consequence, this would free up more O to form water ice, making the Pluto-Charon system underdense compared to the observed value. By employing an updated set of solar abundances derived from solar neutrinos and solar wind measurements, we obtain a higher C/O which is consistent with the observed Pluto-Charon system value, which has been argued as the most representative value we have for the bulk composition of large KBOs (Barr \& Schwamb 2016, Bierson \& Nimmo 2019, McKinnon et al. 2021). Within the range of uncertainties, these values also generally overlap with a recent photospheric determination of heavy element abundances from the 3D, non-LTE radiative transfer model by Magg et al. 2022. An increase in the abundances of heavy elements may support interpretations from helioseismology. Furthermore, from these revised abundances we estimate the refractory-to-ice mass ratio in cometary nuclei to be $\sim$ $1.9-2.5$, which is within the plausible range from Rosetta. A future comet sample return mission could test this prediction. In addition, we test this new dataset with compositional data from CI chondrites, considering that analyses of Ryugu and Bennu samples both show an apparent absence of sulfates (Nakamura et al. 2023, Yokoyma et al. 2023, King et al. 2024). We show that a lack of sulfates significantly reduces the inferred quantity of water ice accreted into CI chondrite parent bodies. This is consistent with the predicted amount of water ice in planetesimals with the carbon speciation of CI chondrites based on the new dataset of solar abundances.\\

As the protosolar abundances control the chemistry of the protosolar nebula and the composition of planetary building blocks, this revision will have significant implications to understand heavy element abundances in the atmospheres of giant planets and the links to their origin and evolution (Helled \& Lunine 2014; Teanby et al. 2020; Mousis et al. 2021). For example, the global water abundances in Jupiter’s atmospheric envelope remains uncertain with crucial implications for Jupiter’s origin (Helled et al. 2022). While interior models and thermochemical modelling generally favor a low water abundance in the envelope (Helled et al. 2022, Cavalié et al. 2023), Juno’s measurements at a single latitude near the equator suggest super-solar enrichments (Li et al. 2020, 2024). Based on a broad set of solar and cosmochemical data, our proposed increase in the solar carbon-to-oxygen C/O ratio implies that planetesimals, may contain less water ice, which supports the possibility that water is depleted in Jupiter’s interior, or at least less enriched relative to solar than is carbon. \\

Our predictions can be tested through several future efforts: 1) Upcoming solar neutrino experiments, like those at the SNO+ (Sudbury Neutrino Observatory) and the Jiangmen Underground Neutrino Observatory – JUNO (Conroy 2024), along with helioseismology models; 2) Current analyses of Bennu samples, which we anticipate finding a higher water content and/or greater porosity compared with CI chondrites; 3) Future comet sample return missions, which should reveal a refractory-to-ice mass ratio of approximately $1.9-2.5$ on comets, and potentially provide precise values for the oxygen isotope ratios of primordial water, particularly if the targets of these missions include comet Hartley 2. The foreseeable future holds great promise for new models and data to provide a more coherent understanding of the Sun's internal composition. As the Sun’s heavy element abundances serve as benchmarks for understanding other stars’ elemental compositions, this will have significant implications to understand the formation and evolution of other stars and planetary systems, and even further, to gain a broader perspective of galactic chemical evolution.

\begin{acknowledgments}
\textbf{Acknowledgment}: N.T thanks Conel Alexander, Scott Bolton, Rob Ebert, Kelly Miller, Olivier Mousis, Charity Phillips-Lander, Danna Qasim, Darryl Seligman, Xinting Yu and Juno’s Origins Working Group for various discussions, which inspired this current work. N.T. also acknowledges J. Tran Thanh Van and Rencontres du Vietnam for the early introduction and inspiration to neutrino and particle physics. N.T and C.R.G acknowledge funding support from SwRI’s Internal Research \& Development program (grant 15-R6321) and the Heising-Simons Foundation (grant 2023–4657). We thank the reviewer for helpful suggestions which improved the paper.
\end{acknowledgments}


\newpage
\begin{appendix}

\section{Influence of C/O ratio on the density of KBOs}
Similar to the calculations presented in Sections 2 and 3.2, here we investigate the influence of different solar carbon-to-oxygen (C/O) ratios on the predicted density of KBOs. Fig A1 and A2 show the two endmember scenarios, in which the corresponding C/O values (0.63, Fig A1 and 0.83, Fig A2) touch the lower bound and upper bound limits for the density of large KBOs. This constraint enables refinement of the estimated solar C abundance and the solar C/O from the solar neutrino measurements and solar wind data (Fig 3, main text).\\

\renewcommand{\thefigure}{A\arabic{figure}}
\setcounter{figure}{0}
\begin{figure}[ht!]
\plotone{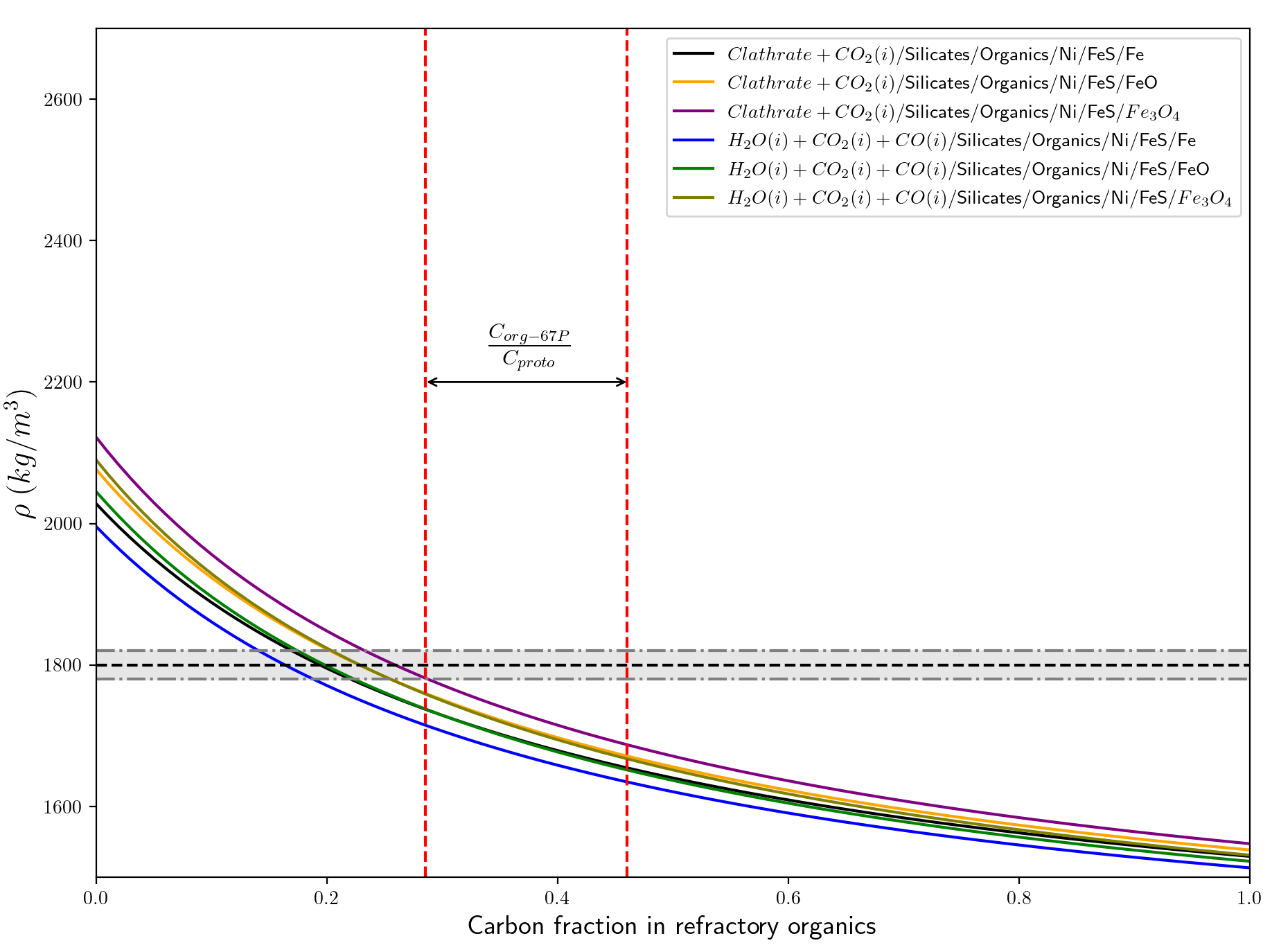}
\caption{The predicted uncompressed density of the Pluto-Charon system (colored curves) as the fraction of the protosolar C inventory contained in refractory organics increases from 0\% to 100\% under a C/O $\sim$ 0.63. The horizontal black dashed line represents the average uncompressed density of the system $\sim$ 1800 kg $m^{-3}$ (Barr \& Schwamb 2016; McKinnon et al. 2017), taken to be the most representative value we have for the bulk composition of large KBOs. The gray area shows the 3$\sigma$ uncertainty ($\pm$ 20 kg $m^{-3}$, McKinnon et al. 2017, Nimmo et al. 2017, Brozović et al. 2015).}  
\label{fig:A1}
\end{figure}

\newpage
\begin{figure}[ht!]
\plotone{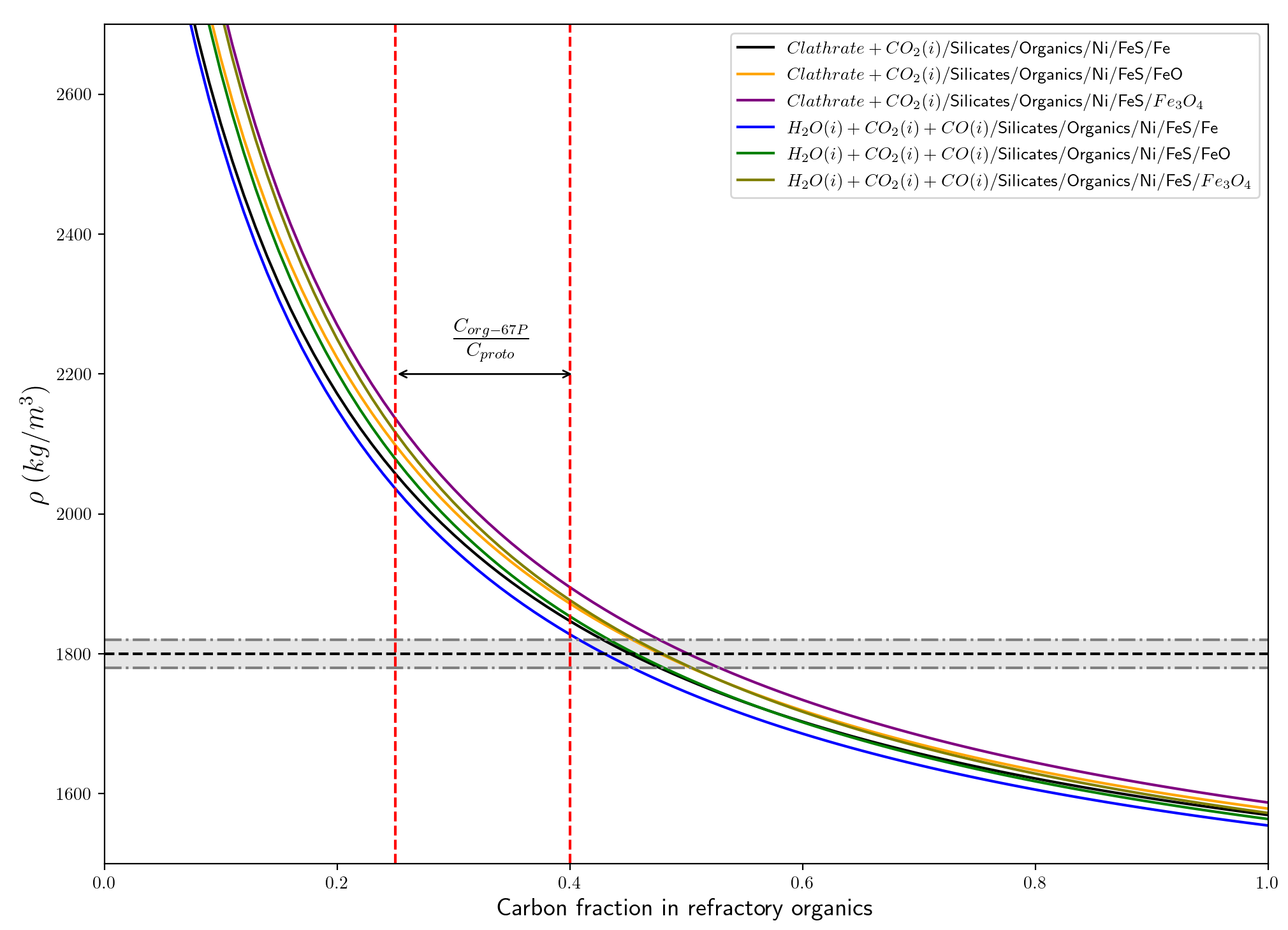}
\caption{The predicted uncompressed density of the Pluto-Charon system (colored curves) as the fraction of the protosolar C inventory contained in refractory organics increases from 0\% to 100\% under a C/O $\sim$ 0.83. The horizontal black dashed line represents the average uncompressed density of the system $\sim$ 1800 kg $m^{-3}$ (Barr \& Schwamb 2016; McKinnon et al. 2017), taken to be the most representative value we have for the bulk composition of large KBOs. The gray area shows the 3$\sigma$ uncertainty ($\pm$ 20 kg $m^{-3}$, McKinnon et al. 2017, Nimmo et al. 2017, Brozović et al. 2015).}  
\label{fig:A2}
\end{figure}

\newpage
\section{Predicted refractory-to-ice mass ratio in comets}
\renewcommand{\thefigure}
{B\arabic{figure}}
\setcounter{figure}{0}
\begin{figure}[ht!]
\plotone{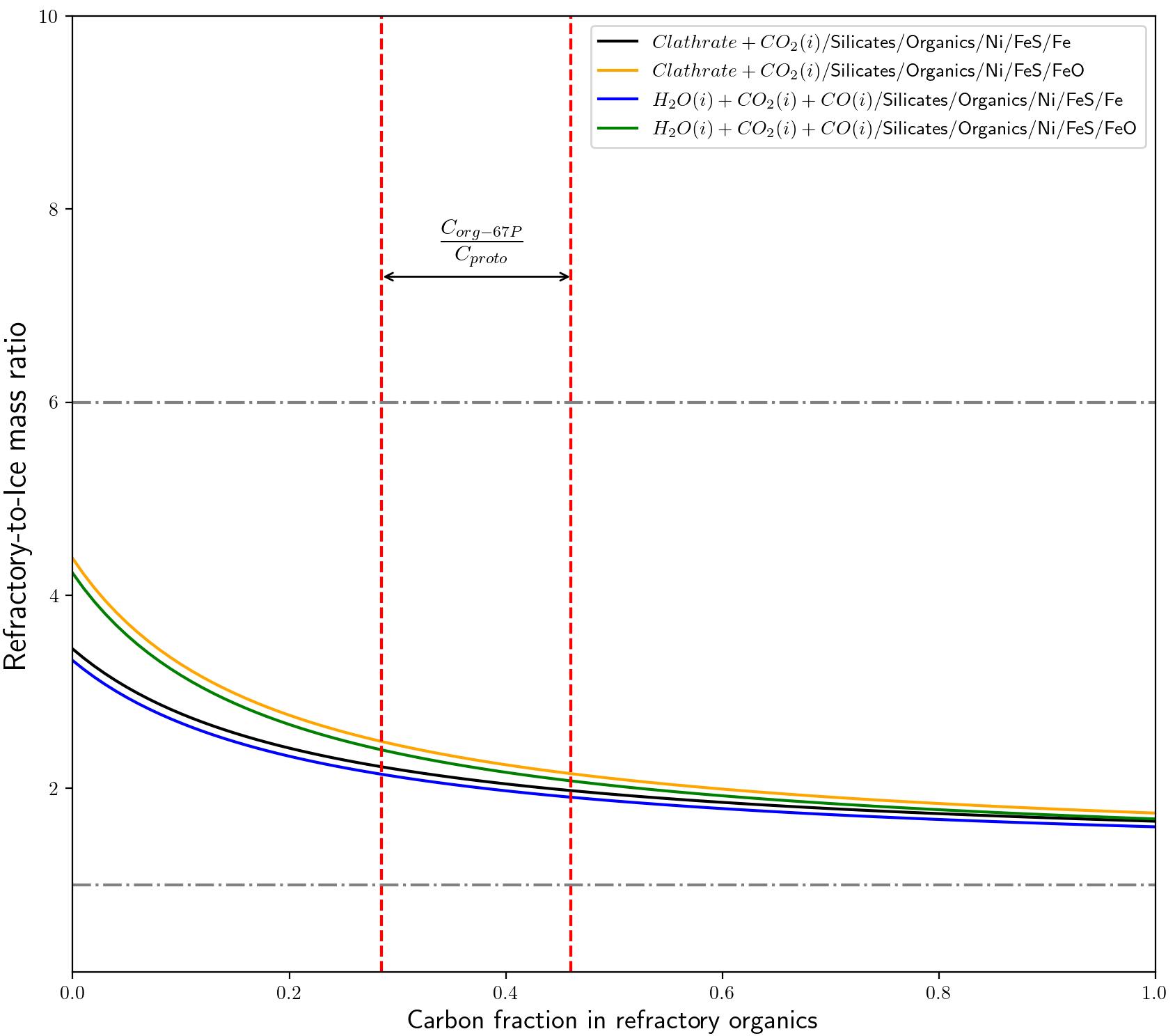}
\caption{The predicted refractory-to-ice mass ratio in cometary nuclei (color curves) fall within the plausible ranges derived from Rosetta (R/I $\sim$ 1-6, between the gray dash-dot lines) (Choukroun et al. 2020). While a lower or higher value is possible, this range encompasses most values derived from different instruments on Rosetta. We also note that comet 67P is an evolved comet, and other processes (i.e., redistribution and deposition of dust from the more illuminated hemisphere to the other hemisphere, Thomas et al. 2015) may complicate the interpretation. }  
\label{fig:B1}
\end{figure}

\newpage
\section{Ryugu bulk abundances vs CI chondrites}
Here, we compile the latest data on the average composition of Ryugu samples for 67 elements (Naraoka et al. 2023, Yokoyama et al. 2023) and compare it with data for CI chondrites from Lodders 2021. The full dataset is available at: \url{ https://doi.org/10.5281/zenodo.13345297}. Among these data, the effects of space weathering on surface and subsurface samples are also evaluated.\\

Surface samples stored in Chamber A were collected during the first touchdown on Ryugu’s surface on February 21, 2019. Samples in Chamber C were collected near an artificial crater (diameter of $\sim$ 14 m) on Ryugu’s surface. They are possibly ejecta from the north side of the crater by the second touchdown on July 11, 2019 (Tsuda et al. 2020). Ryugu's surface is bombarded by cosmic rays, UV radiation (Nishiizumi et al. 2022), and experiences thermal cycling during diurnal and seasonal illumination cycles (Kitazato et al. 2021). The ${}^{10}_{}Be$ concentrations in all Chamber C samples are the same but are lower than those of the Chamber A samples, suggesting that they were shielded at a depth of 0.7-1.3 m and were ejected from the lower portions of the crater (Nishiizumi et al. 2022).
\renewcommand{\thefigure}
{C\arabic{figure}}
\setcounter{figure}{0}
\begin{figure}[ht!]
\plotone{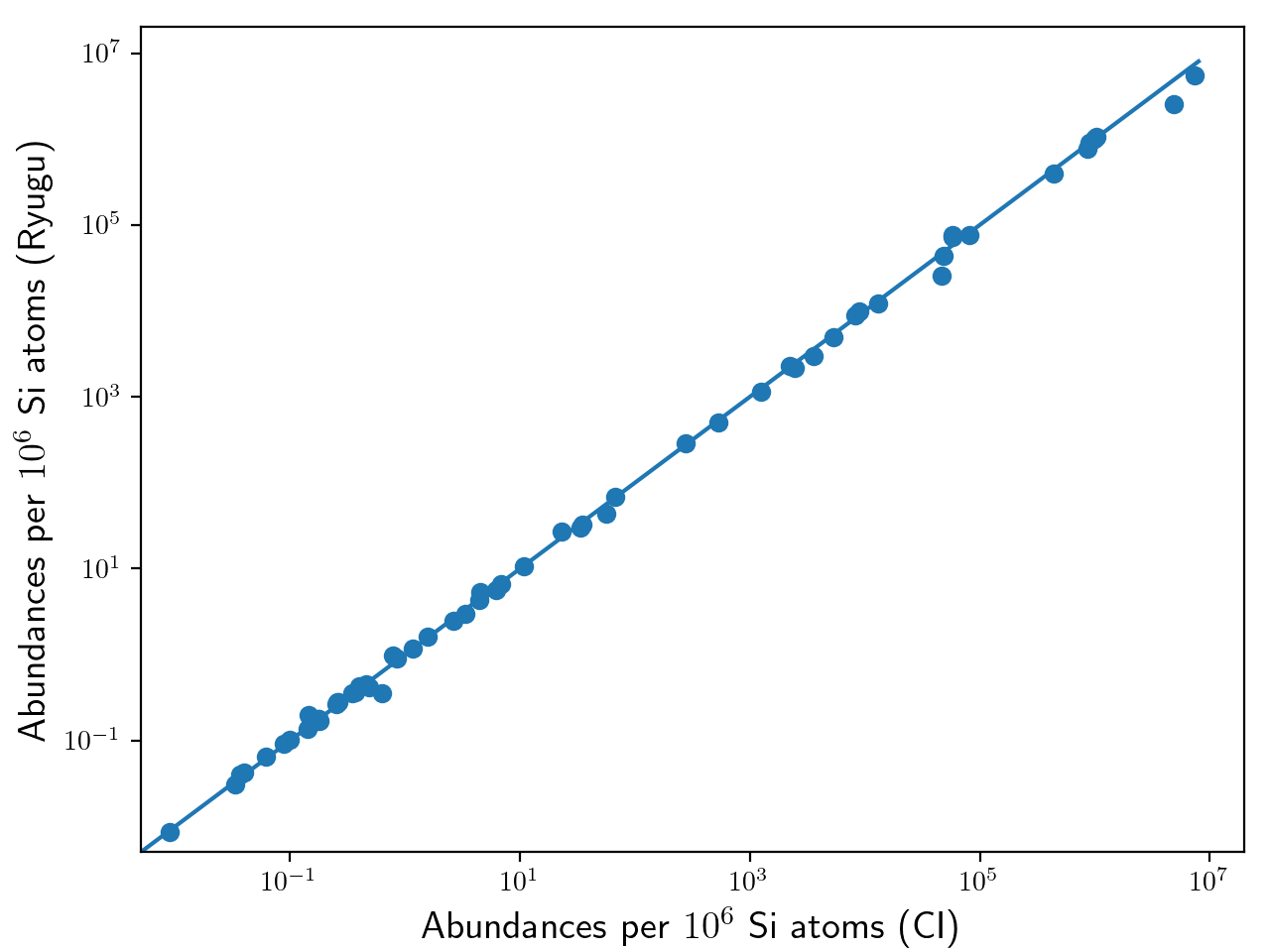}
\caption{The bulk abundance of 67 elements in Ryugu samples compared with data for CI chondrites from Lodders 2021. The full dataset is available at: \url{ https://doi.org/10.5281/zenodo.13345297}}
\label{fig:C1}
\end{figure}

Given these considerations, our compiled data mostly consists of analyzed results from the subsurface samples, except for elements where subsurface data have not yet been reported in the literature. Fig. C1 shows the  average abundances of elements per $10^6$ Si atoms in Ryugu samples and CI chondrites (Lodders 2021). The Ryugu samples are mostly similar to CI chondrites. Ryugu samples are more chemically pristine than CI chondrites as they have not  experienced terrestrial contamination, including the alteration of organics and phyllosilicate structures, the adsorption of terrestrial water, and the formation of sulfates and ferric hydroxides (Nakamura et al. 2023, Yokoyama et al. 2023, Takano et al. 2024). However, due to a combination of impact heating, solar heating, solar wind irradiation, and long-term exposure to the ultrahigh vacuum of space, the interlayer water in Ryugu's phyllosilicates may have been partially lost to space (Yokoyama et al. 2023). These factors likely contribute to the differences in H, N, and O between Ryugu samples and CI chondrites. \\

\newpage
\section{The initial oxygen isotope abundances in CI chondrite water}

In this section, we use the classic oxygen isotopes exchange model presented in Clayton \& Mayeda 1999. In that model, oxygen isotopes can be exchanged during the hydration reaction of olivine and pyroxene to serpentine written as: $Mg_2SiO_4 + MgSiO_3 + 2 H_2O \rightarrow Mg_3Si_2O_5(OH)_4$. Given $x$: mole of O in water per mole of O in initial rock, $f$: fraction of initial rock altered, $\delta_{w}^{i}$ and $\delta_{w}^{f}$ are the initial and final isotopic compositions of water, $\delta_{r}^{i}$ is the initial isotopic composition of rock (olivine and pyroxene), and $\delta_{s}^{f}$ is the final isotopic composition of rock (serpentine), the isotopic mass balance equation is:\\

\textbf{[C1]}: $7x\delta_{w}^{i} + 7\delta_{r}^{i} = (7x-2f)\delta_{w}^{f} + 7(1-f)\delta_{r}^{i} + 9f\delta_{s}^{f} $
\\

Our predicted water-to-accreted rock mass ratio ($W/R$) can be converted to $x$ (moles of O in water/mole of O in initial rock) as:\\

\textbf{[C2]}: $\frac{m_{w}}{m_{r}} = \frac{\mu_{w}n_{w}}{\mu_{r}n_{r}}$ = $\frac{3.5x\mu_{w}}{\mu_{r}}$
\\

The factor 3.5 appears as there are 3.5 moles of oxygen per 1 mole of the simplified starting rock. Here, $\mu_w$ = 18 g $mol^{-1}$, $\mu_{Mg_{1.5}SiO_{3.5}}$ = 120 g $mol^{-1}$.\\

Rearranging Equation [C1], the initial isotopic composition of water can be calculated using:\\

\textbf{[C3]}: $\delta_{w}^{i} = \delta_{w}^{f} + (\frac{f}{7x})(9 \delta_{s}^{f} - 2\delta_{w}^{f} -7\delta_{r}^{i}).$\\

As discussed in Section 3.2, it is challenging to determine the isotopic composition of anhydrous olivine and pyroxene for CI chondrites, as they show large grain-to-grain variations and follow a bimodal distribution (Leshin et al. 1997, Kawasaki et al. 2022). Despite this, a common procedure is to adopt an average value for the initial isotopic composition of the anhydrous grains: $\delta_{r}^{18}$ = 4.8 {\textperthousand} and $\delta_{r}^{17}$ = 1.8 {\textperthousand} (Leshin et al. 1997, Clayton \& Mayeda 1999). We adopt the rest of the input data as measured in Ryugu samples, for serpentine: $\delta_{s}^{18}$ = 18.6 {\textperthousand} and $\delta_{s}^{17}$ = 9.2 {\textperthousand}, and for the final water that equilibrated with altered rock: $\delta_{w}^{18}$ = 1.0 {\textperthousand} and $\delta_{w}^{17}$ = 0.3 {\textperthousand} (Yokoyama et al. 2023). As almost all of the initial rock was altered in CI chondrites, we assume $f$ = 1. Our model's lower and upper bounds for the water-to-rock mass ratio ratio can be calculated from the initial oxygen accreted as water ice in Table 3; this gives $W/R$ = $0.23-0.30$ and the corresponding $x$ = $0.43–0.57$. Thus, the initial isotopic compositions of water is deduced to have been: $\delta_{w}^{18}$ = $34.0-44.8$ {\textperthousand} and $\delta_{w}^{17}$ = $17.7-23.4$ {\textperthousand}. These constraints satisfy the lower limit values derived by Clayton \& Mayeda 1999 ($\delta_{w}^{18}$ $\geq$ 24.1 {\textperthousand}, $\delta_{w}^{17}$ $\geq$ 15.2 {\textperthousand}), who attempted to fit a single isotopic composition of a water reservoir to three separate carbonaceous chondrite groups: CI, CM and CR.

\end{appendix}

\newpage
\section*{REFERENCES}

Agnor, C. B., \& Hamilton, D. P. 2006, Nature, 441, 192\\
Agostini, M., Altenmüller, K., Appel, S., et al. 2020a, Nature, 587, 577\\
Agostini, M., Altenmüller, K., Appel, S., et al. 2020b, Eur Phys J C, 80, 1091\\
Airieau, S. A., Farquhar, J., Thiemens, M. H., et al. 2005, GeCoA, 69, 4167\\
Alexander, C. M. O., Fogel, M., Yabuta, H., \& Cody, G. D. 2007, GeCoA, 71, 4380\\
Alexander, C. M. O., Bowden, R., Fogel, M. L., et al. 2012, Science, 337, 721\\
Alexander, C. M. O., Bowden, R., Fogel, M. L., \& Howard, K. T. 2015, Meteorit. Planet. Sci., 50, 810\\
Alexander, C. M. O., Cody, G. D., De Gregorio, B. T., Nittler, L. R., \& Stroud, R. M. 2017, Geochemistry, 77, 227\\
Alexander, C. M. O. 2019, GeCoA, 254, 277\\
Allende Prieto, C. 2016, Living Rev Sol Phys, 13, 1\\
Anders, E., \& Grevesse, N. 1989, GeCoA, 53, 197\\
Appel, S., Bagdasarian, Z., et al. 2022, Phys Rev Lett, 129, 252701\\
Asplund, M., Grevesse, N., \& Sauval, A. J. 2005, 336, 25\\
Asplund, M., Grevesse, N., Sauval, A. J., \& Scott, P. 2009, Annu. Rev. Astron. Astrophys., 47, 481\\
Asplund, M., Amarsi, A. M., \& Grevesse, N. 2021, A\&A, 653, A141\\
Bahcall, J. N., Basu, S., Pinsonneault, M., \& Serenelli, A. M. 2005, ApJ, 618, 1049\\
Bahcall, J. N., Serenelli, A. M., \& Basu, S. 2006, ApJS, 165, 400\\
Bailey, J. E., Nagayama, T., Loisel, G. P., et al. 2015, Nature, 517, 56\\
Bardyn, A., Baklouti, D., Cottin, H., et al. 2017, MNRAS, 469, S712\\
Barr, A. C., \& Schwamb, M. E. 2016, MNRAS, 460, 1542\\
Basilico, D., Bellini, G., et al. 2023, Phys Rev D, 108, 102005\\
Basu, S., \& Antia, H. M. 2004, ApJ, 606, L85\\
Basu, S., \& Antia, H. M. 2008, Phys. Rep., 457, 217\\
Bethe, H. A. 1939, Phys Rev, 55, 434\\
Bierson, C. J., \& Nimmo, F. 2019, Icarus, 326, 10\\
Bochsler, P. 2007, A\&A, 471, 315\\
Bouquet, A., Miller, K. E., Glein, C. R., \& Mousis, O. 2021, A\&A, 653, A59\\
Brearley, A. J. 2006, in Meteorites and the Early Solar System II, ed. Lauretta, D.S, McSween, H. Y. (Tucson, AZ: Univ. Arizona Press)\\
Brown, M. E., Barkume, K. M., Ragozzine, D., \& Schaller, E. L. 2007, Nature, 446, 294\\
Brozović, M., Showalter, M. R., Jacobson, R. A., \& Buie, M. W. 2015, Icarus, 246, 317\\
Buldgen, G., Eggenberger, P., Noels, A., et al. 2023, A\&A, 669, L9\\
Cameron, A. G. W. 1973, Space Sci Rev, 15, 121\\
Cameron, A. G. W. 1982, Elemental and Nuclidic Abundances in the Solar System, https://ui.adsabs.harvard.edu/abs/1982ena..conf...23C\\
Cañas, M. H., Lyra, W., Carrera, D., et al. 2024, Planet. Sci. J, 5, 55\\
Canup, R. M., \& Ward, W. R. 2002, AJ, 124, 340\\
Canup, R. M. 2005, Science, 307, 546\\
Canup, R. M., \& Ward, W. R. 2006, Nature, 441, 834\\
Cavalié, T., Lunine, J., \& Mousis, O. 2023, Nat Astron, 7, 678\\
Canup, R. M., Kratter, K. M., \& Neveu, M. 2021, in: The Pluto System after New Horizons, ed. Stern, A. et al. (Tucson, AZ: Univ. Arizona Press) \\
Choukroun, M., Altwegg, K., Kührt, E., et al. 2020, Space Sci Rev, 216, 44\\
Christensen-Dalsgaard, J., Mauro, M. P. D., Houdek, G., \& Pijpers, F. 2009, A\&A, 494, 205\\
Clayton, R. N., \& Mayeda, T. K. 1999, GeCoA, 63, 2089\\
Conroy, G. 2024, Nature, 627, 715\\
Ćuk, M., Dones, L., \& Nesvorný, D. 2016, ApJ, 820, 97\\
Desch, S. J., Kalyaan, A., \& Alexander, C. M. O. 2018, ApJS, 238, 11\\
Emery, J. P., Wong, I., Brunetto, R., et al. 2024, Icarus, 414, 116017\\
Fard, K. B., \& Smith, I. B. 2024, Icarus, 410, 115895\\
Fegley, B. 2000, Space Sci. Rev., 92, 177\\
Fray, N., Bardyn, A., Cottin, H., et al. 2016, Nature, 538, 72\\
Fujiya, W. 2018, EPSL, 481, 264\\
Glein, C. R., \& Waite, J. H. 2018, Icarus, 313, 79\\
Goldberg, L., Muller, E. A., \& Aller, L. H. 1960, ApJS, 5, 1\\
Goldreich, P., Murray, N., Longaretti, P. Y., \& Banfield, D. 1989, Science, 245, 500\\
Gounelle, M., \& Zolensky, M. E. 2001, Meteorit. Planet. Sci., 36, 1321\\
Gounelle, M., \& Zolensky, M. E. 2014, Meteorit. Planet. Sci., 49, 10\\
Greenberg, J. M. 1998, A\&A, 330\\
Grevesse, N., \& Sauval, A. J. 1998, Space. Sci. Rev, 85, 161\\
Grundy, W. M., Wong, I., Glein, C. R., et al. 2024, Icarus, 411, 115923\\
Harrington-Pinto, O. H., Womack, M., Fernandez, Y., \& Bauer, J. 2022, Planet Sci J, 3, 247\\
Hartogh, P., Lis, D. C., Bockelée-Morvan, D., et al. 2011, Nature, 478, 218\\
Haxton, W. C., \& Serenelli, A. M. 2008, ApJ, 687, 678\\
Haxton, W. C., Robertson, R. G. H., \& Serenelli, A. M. 2013, Annu. Rev. Astron. Astrophys., 21\\
Hayes, A. G., Jr., Nakamura-Messenger, K., Squyres, S. W., et al. 2020, AGU 2020, P089\\
Heber, V. S., McKeegan, K. D., Steele, R. C. J., et al. 2021, ApJ, 907, 15\\
Helled, R., Lunine, J. I. 2014, MNRAS, 441, 3 \\
Helled, R., Stevenson, D. J., Lunine, J. I., et al. 2022, Icarus, 378, 114937\\
Hinkel, N. R., Timmes, F. X., Young, P. A., Pagano, M. D., \& Turnbull, M. C. 2014, AJ, 148, 54\\
Hoarty, D. J., Morton, J., Rougier, J. C., et al. 2023, Phys. Plasmas, 30, 063302 \\
Huss, G. R., Koeman-Shields, E., Jurewicz, A. J. G., et al. 2020, Meteorit. Planet. Sci., 55, 326\\
Isnard, R., Bardyn, A., Fray, N., et al. 2019, A\&A, 630, A27\\
Jessberger, E. K., Christoforidis, A., \& Kissel, J. 1988, Nature, 332, 691\\
Jofré, P., Heiter, U., \& Soubiran, C. 2019, Annu. Rev. Astron. Astrophys. 57, 571\\
Johnson, T. V., \& Lunine, J. I. 2005, Nature, 435, 69\\
Johnson, T. V., Mousis, O., Lunine, J. I., \& Madhusudhan, N. 2012, ApJ, 757, 192\\
Kawasaki, N., Nagashima, K., Sakamoto, N., et al. 2022, Sci. Adv., 8, 50\\
King, A. J., Tu, V., Schofield, P. F., et al. 2024, 55th LPSC, The Woodlands TX, 1109\\
Kiss, C., Marton, G., Parker, A. H., et al. 2019, Icarus, 334, 3\\
Kissel, J., Brownlee, D. E., Büchler, K., et al. 1986, Nature, 321, 336\\
Kissel, J., \& Krueger, F. R. 1987, Nature, 326, 755\\
Kitazato, K. et al. 2021, Nat. Astron, 5\\
Krijt, S., Bosman, A. D., Zhang, K., et al. 2020, ApJ, 899, 134\\
Kruijer, T. S., Burkhardt, C., Budde, G., \& Kleine, T. 2017, PNAS, 114, 6712\\
Kunitomo, M., \& Guillot, T. 2021, A\&A, 655, A51\\
Kunitomo, M., Guillot, T., \& Buldgen, G. 2022, A\&A, 667, L2\\
Laming, J. M., Heber, V. S., Burnett, D. S., et al. 2017, ApJ, 851, L12 \\
Lauretta, D. S., Connolly, H., Aebersold, J. E., et al. 2024, Meteorit. Planet. Sci., doi: 10.1111/maps.14227\\
Leshin, L. A., Rubin, A. E., \& McKeegan, K. D. 1997, GeCoA, 61, 835\\
Levison, H. F., \& Duncan, M. J. 1997, Icarus, 127, 13\\
Li, C., Ingersoll, A., Bolton, S., et al. 2020, Nat Astron, 4, 609\\
Li, C., Allison, M., Atreya, S., et al. 2024, Icarus, 414, 116028\\
Lodders, K. 2003, ApJ, 591, 1220\\
Lodders, K. 2020, in: Oxford Research Encyclopedia of Planetary Science (Oxford Univ. Press)\\
Lodders, K. 2021, Space Sci Rev, 217, 44\\
Luna, R., Millán, C., Domingo, M., Santonja, C., \& Satorre, M. Á. 2022, ApJ, 935, 134\\
Lunine, J. I., \& Stevenson, D. J. 1982, Icarus, 52, 14\\
Macke, R. J., Consolmagno, G. J., \& Britt, D. T. 2011, Meteorit. Planet. Sci., 46, 1842\\
Magg, E., Bergemann, M., Serenelli, A., et al. 2022, A\&A, 661, A140\\
McKinnon, W. B., \& Mueller, S. 1988, Nature, 335, 240\\
McKinnon, W. B., \& Leith, A. C. 1995, Icarus, 118, 392\\
McKinnon, W. B., Lunine, J.I., Banfield, D. 1995, in: Neptune and Triton, ed. Cruikshank, D. P. et al. (Tucson, AZ: Univ. Arizona Press)\\
McKinnon, W. B., Stern, S. A., Weaver, H. A., et al. 2017, Icarus, 287, 2\\
McKinnon, W. B., Glein, C. R., Bertrand, T., \& Rhoden, A. R. 2021, in: The Pluto System after New Horizons, ed. Stern, A. et al. (Tucson, AZ: Univ. Arizona Press)\\
Miller, K. E., Glein, C. R., \& Waite, J. H. 2019, ApJ, 871, 59\\
Mosqueira, I., \& Estrada, P. R. 2003, Icarus, 163, 198\\
Mousis, O., Lunine, J. I., \& Aguichine, A. 2021, ApJL, 918, L23\\
Mousis, O., Schneeberger, A., Lunine, J. I., et al. 2023, ApJL, 944, L37\\
Nagayama, T., Bailey, J. E., Loisel, G. P., et al. 2019, Phys Rev Lett, 122, 235001\\
Nakamura, E., Kobayashi, K., Tanaka, R., et al. 2022, PJA-B, 98, 227\\
Nakamura, T., Matsumoto, M., Amano, K., et al. 2023, Science, 379, eabn8671\\
Naraoka, H. et al. 2023, Science, 379, eabn9033\\
Néri, A., Guyot, F., Reynard, B., \& Sotin, C. 2020, EPSL, 530, 115920\\
Nesvorný, D., Vokrouhlický, D., Dones, L., et al. 2017, ApJ, 845, 27\\
Nieva, M.-F., \& Przybilla, N. 2012, A\&A, 539, A143\\
Nimmo, F., Umurhan, O., Lisse, C. M., et al. 2017, Icarus, 287, 12\\
Nishiizumi, K. et al. 2022, 53 LPSC, The Woodlands TX, 1777\\
Öberg, K. I., Murray-Clay, R., \& Bergin, E. A. 2011, ApJ, 743, L16\\
Ortiz, J. L., Santos-Sanz, P., Sicardy, B., et al. 2017, Nature, 550, 219\\
Owen, T., Mahaffy, P., Niemann, H. B., et al. 1999, Nature, 402, 269\\
Palme, H. et al. 2014, Treatise on Geochemistry, Elsevier, 2\\
Pätzold, M., Andert, T. P., Hahn, M., et al. 2019, MNRAS, 483, 2337\\
Pekmezci, G. S., Johnson, T. V., Lunine, J. I., \& Mousis, O. 2019, ApJ, 887, 3\\
Piani, L., Nagashima, K., Kawasaki, N., et al. 2023, ApJL, 946, L43\\
Pilleri, P., Reisenfeld, D. B., Zurbuchen, T. H., et al. 2015, ApJ, 812, 1\\
Podolak, M., Pollack, J. B., \& Reynolds, R. T. 1988, Icarus, 73, 163\\
Pollack, J. B., Hollenbach, D., Beckwith, S., et al. 1994, ApJ, 421, 615\\
Prinn, R. G. P., \& Fegley, B., Jr. 1989, ed. Atreya, S. K. et al., in: Origin and evolution of planetary and satellite atmospheres (Tucson, AZ: Univ. Arizona Press)  \\
Reynard, B., \& Sotin, C. 2023, EPSL, 612, 118172\\
Russell, H. N. 1929, ApJ, 70, 11\\
Serenelli, A. M., Basu, S., Ferguson, J. W., \& Asplund, M. 2009, ApJ, 705, L123\\
Serenelli, A. M., \& Basu, S. 2010, ApJ, 719, 865\\
Serenelli, A. M., Haxton, W. C., \& Peña-Garay, C. 2011, ApJ, 743, 24\\
Serenelli, A., Peña-Garay, C., \& Haxton, W. C. 2013, Phys Rev D, 87, 043001\\
Simonelli, D. P., Pollack, J. B., Mckay, C. P., Reynolds, R. T., \& Summers, A. L. 1989, Icarus, 82, 1\\
Sloan, E. D., Koh, C. A. 2007, Clathrate Hydrates of Natural Gases (3rd ed.; Boca Raton: CRC Press)\\
von Steiger, R., \& Zurbuchen, T. H. 2016, ApJ, 816, 13\\
Takano, Y. et al. 2024, Nat. Commun, 15, 5708\\
The SNO+ collaboration, T. S., Albanese, V., Alves, R., et al. 2021, J Inst, 16, P08059\\
Teanby, N. A., Irwin, P. G. J., Moses, J. I., \& Helled, R. 2020, Philos. Trans. R. Soc. A, 378 , 20190489\\
Thomas, N., Davidsson, B., El-Maarry, M. R., et al. 2015, A\&A, 583, A17\\
Tsuchiyama, A., Matsumoto, M., Matsuno, J., et al. 2024, GeCoA, 375, 146\\
Tsuda, Y. et al. 2020, Acta Astronaut, 171\\
Villante, F. L., \& Serenelli, A. 2021, Front Astron Space Sci, 7\\
Vinyoles, N., Serenelli, A. M., Villante, F. L., et al. 2017, ApJ, 835, 202\\
Volk, K., \& Malhotra, R. 2008, ApJ, 687, 714\\
Waite, J. H., Glein, C. R., Perryman, R. S., et al. 2017, Science, 356, 155\\
Wang, H., Weiss, B. P., Bai, X.-N., et al. 2017, Science, 355, 623\\
Wiens, R. C., Burnett, D. S., Neugebauer, M., \& Pepin, R. O. 1991, Geophys. Res. Lett, 18, 207\\
Wong, M. H., Lunine, J. I., Atreya, S. K., et al. 2008, Rev. Mineralogy. Geochem, 68, 219\\
Yada, T. et al. 2022, Nat. Astron, 6\\
Yamaguchi, A. et al. 2023, Nat. Astron, 7\\
Yoshimura, T. et al. 2023, Nat. Commun, 14, 5284\\
Yokoyama, T., Nagashima, K., Nakai, I., et al. 2023, Science, 379, eabn7850\\
Young, P. R. 2018, ApJ, 855, 15\\
Zolensky et al. 2022, 53 LPSC, The Woodlands TX, 1451





\end{document}